\documentclass[review,number,sort&compress]{elsarticle}
\usepackage{lineno}
\usepackage{amssymb}
\usepackage{amsmath}

\def\lsim{\mathrel{\rlap{\lower3pt\hbox{$\sim$}}
    \raise2pt\hbox{$<$}}}                % less than or approx. symbol

\newcommand {\DF} {\mbox{D-F}}
\newcommand {\FD} {\mbox{F-D}}

\journal{NIM-A}
\begin{document}
\begin{frontmatter}

\title{Performance of the first prototype of the CALICE scintillator strip electromagnetic calorimeter}

\begin{abstract} %NIM
A first prototype of a scintillator strip-based electromagnetic calorimeter was built, 
consisting of 26 layers of tungsten absorber plates interleaved with planes of $45\times10\times3~\mathrm{mm}^3$ plastic scintillator 
strips. % in an alternating configuration (horizontally and vertically) to achieve effective 10$\times$10~mm$^2$ granularity.
Data were collected using a positron test beam at DESY with momenta between 1 and 6 GeV/c.
The prototype's performance is presented in terms of the linearity and resolution of the energy measurement.
These results represent an important milestone in the development of highly granular calorimeters using scintillator strip 
technology.
% added dj 12/2/2014
A number of possible design improvements were identified, which should be implemented in a future detector of this type.
This technology is being developed for a future linear collider experiment, 
aiming at the precise measurement of jet energies using particle flow techniques.
\end{abstract} %NIM

\author{%\centering 
{\large \bf The CALICE Collaboration}
}
\address{\ }

%%Nigel Watson 20131108 \author{\centering
%%Nigel Watson 20131108 C.\,Adloff, 
%%Nigel Watson 20131108 J.-J.\,Blaising, 
%%Nigel Watson 20131108 M.\,Chefdeville, 
%%Nigel Watson 20131108 C.\,Drancourt,
%%Nigel Watson 20131108 R.\,Gaglione, 
%%Nigel Watson 20131108 N.\,Geffroy, 
%%Nigel Watson 20131108 Y.\,Karyotakis, 
%%Nigel Watson 20131108 I.\,Koletsou, 
%%Nigel Watson 20131108 J.\,Prast,
%%Nigel Watson 20131108 G.\,Vouters 
%%Nigel Watson 20131108 \\ \it
%%Nigel Watson 20131108 Laboratoire d'Annecy-le-Vieux de Physique des Particules, Universit\'{e} de Savoie,
%%Nigel Watson 20131108 CNRS/IN2P3,
%%Nigel Watson 20131108 9 Chemin de Bellevue BP110, F-74941 Annecy-le-Vieux CEDEX, France
%%Nigel Watson 20131108 }

\author{
K.\,Francis$^a$,
J.\,Repond, 
J.\,Schlereth, 
J.\,Smith$^b$,
 %%\endnote{Also at University of Texas, Arlington},
L.\,Xia
}
\address{
Argonne National Laboratory,
9700 S.\ Cass Avenue,
Argonne, IL 60439-4815,
USA}

\author{
E.\,Baldolemar, 
J.\,Li$^c$, %%\endnote{Deceased}, 
S.\,T.\,Park, 
M.\,Sosebee, 
A.\,P.\,White, 
J.\,Yu
}\address{
Department of Physics, SH108, University of Texas, Arlington, TX 76019, USA
}

\author{
G.\,Eigen 
}\address{
University of Bergen, Inst.\,of Physics, Allegaten 55, N-5007 Bergen, Norway
}

\author{
Y.\,Mikami, 
N.\,K.\,Watson 
}\address{
University of Birmingham,
School of Physics and Astronomy,\\
Edgbaston, Birmingham B15 2TT, UK
}

\author{
% G.\,Mavromanolakis$^c$, %%\endnote{Now at CERN},
M.\,A.\,Thomson, 
D.\,R.\,Ward 
% W.\,Yan$^d$ %{Now at Dept.\ of Modern Physics, Univ.\ of Science and
% Technology of China, 96 Jinzhai Road, Hefei, Anhui, 230026, P.\, R.\, China}
}\address{
University of Cambridge, Cavendish Laboratory, J J Thomson Avenue, CB3 0HE, UK
}

\author{%\centering 
D.\,Benchekroun, 
A.\,Hoummada, 
Y.\,Khoulaki
}\address{
Universit\'{e} Hassan II A\"{\i}n Chock, Facult\'{e} des sciences,\\ B.P. 5366 Maarif, Casablanca, Morocco
}

\author{%\centering 
J.\,Apostolakis, 
%%Nigel Watson 20131108 D.\,Dannheim, 
A.\,Dotti, 
G.\,Folger, 
V.\,Ivantchenko, 
%%Nigel Watson 20131108 W.\,Klempt, 
%%Nigel Watson 20131108 E.\,van der Kraaij$^e$, %%,now at University of Bergen 
%%Nigel Watson 20131108 A.\,-I.\,Lucaci-Timoce, 
A.\,Ribon, 
%%Nigel Watson 20131108 D.\,Schlatter, 
V.\,Uzhinskiy
}\address{
CERN, 1211 Gen\`{e}ve 23, Switzerland
}

\author{%\centering
C.\,C\^{a}rloganu, 
P.\,Gay, 
S.\,Manen, 
L.\,Royer
}\address{
Laboratoire de Physique Corpusculaire, Clermont Universit\'e, Universit\'e Blaise Pascal, CNRS/IN2P3, BP 10448, F-63000 Clermont-Ferrand, France
%Clermont Universit\'e, Universit\'e Blaise Pascal, CNRS/IN2P3, LPC, \\ 
%BP 10448, F-63000 Clermont-Ferrand, France
}

\author{%\centering
M.\,Tytgat,
N.\,Zaganidis
}\address{
Ghent University, Department of Physics and Astronomy, \\
Proeftuinstraat 86, B-9000 Gent, Belgium
}

\author{%\centering
G.\,C.\,Blazey,
A.\,Dyshkant, 
J.\,G.\,R.\,Lima, 
V.\,Zutshi
}\address{
NICADD, Northern  Illinois University,
Department of Physics,
DeKalb, IL 60115,
USA
}

\author{%\centering 
J.\,-Y.\,Hostachy, 
L.\,Morin
}\address{
Laboratoire de Physique Subatomique et de Cosmologie - Universit\'{e} Joseph Fourier Grenoble 1 -
CNRS/IN2P3 - Institut Polytechnique de Grenoble,\\
53, rue des Martyrs,
38026 Grenoble CEDEX, France
}

\author{%\centering 
U.\,Cornett, 
D.\,David, 
A.\,Ebrahimi, 
G.\,Falley, 
K.\,Gadow, 
P.\,G\"{o}ttlicher, 
C.\,G\"{u}nter,
O.\,Hartbrich, 
B.\,Hermberg, 
S.\,Karstensen, 
F.\,Krivan,
K.\,Kr\"uger, 
%%S.\,Lu, 
B.\,Lutz, 
S.\,Morozov, 
V.\,Morgunov$^d$, %%\endnote{On leave from ITEP}, 
C.\,Neub\"user,
M.\,Reinecke, 
F.\,Sefkow, 
P.\,Smirnov,
M.\,Terwort
}\address{
DESY, Notkestrasse 85,
D-22603 Hamburg, Germany
}

\author{%\centering  
%%Nigel Watson 20131108 N.\,Feege, 
E.\,Garutti, 
S.\,Laurien, 
S.\,Lu,  %%\endnote{Also at DESY},
I.\,Marchesini$^e$, %%\endnote{Also at DESY},
M.\,Matysek, 
M.\,Ramilli
}\address{
Univ. Hamburg,
Physics Department,
Institut f\"ur Experimentalphysik,\\
Luruper Chaussee 149,
22761 Hamburg, Germany
}

\author{%\centering 
K.\,Briggl, 
P.\,Eckert, 
T.\,Harion, 
H.\,-Ch.\,Schultz-Coulon, 
W.\,Shen, 
R.\,Stamen
}\address{
 University of Heidelberg, Fakult\"at f\"ur Physik und Astronomie,\\
Albert \"Uberle Str. 3-5 2.OG Ost,
D-69120 Heidelberg, Germany
}

\author{%\centering 
B.\,Bilki$^f$, %%\endnote{Also at Argonne National Laboratory},
E.\,Norbeck$^c$, %%\endnote{Deceased}, 
D.\,Northacker,
Y.\,Onel
}\address{
University of Iowa, Dept. of Physics and Astronomy,\\
203 Van Allen Hall, Iowa City, IA 52242-1479, USA
}

\author{%\centering 
G.\,W.\,Wilson
}\address{
University of Kansas, Department of Physics and Astronomy,\\
Malott Hall, 1251 Wescoe Hall Drive, Lawrence, KS 66045-7582, USA
}

\author{%\centering 
K.\,Kawagoe,
Y.\,Sudo,
T.\,Yoshioka
}\address{
Department of Physics, Kyushu University, Fukuoka 812-8581, Japan
}

\author{%\centering 
P.\,D.\,Dauncey
%%Nigel Watson 20131108 A.\,-M.\,Magnan
}\address{
Imperial College London, Blackett Laboratory,
Department of Physics,\\
Prince Consort Road,
London SW7 2AZ, UK 
}

\author{%\centering 
%%Nigel Watson 20131108 V.\,Bartsch$^i$, %%\endnote{Now at University of Sussex, Physics and Astronomy Department, Brighton, Sussex, BN1 9QH, UK}, 
M.\,Wing
}\address{
Department of Physics and Astronomy, University College London,\\
Gower Street,
London WC1E 6BT, UK
}

\author{%\centering 
F.\,Salvatore$^g$ %%\endnotemark[8]
}\address{
Royal Holloway University of London,
Department of Physics,\\
Egham, Surrey TW20 0EX, UK
}

\author{%\centering 
E.\,Cortina Gil, 
S.\,Mannai
}\address{
Center for Cosmology, Particle Physics and Cosmology (CP3), \\
Universit\'{e} catholique de Louvain, Chemin du cyclotron 2,
1320 Louvain-la-Neuve, Belgium
}

\author{%\centering 
 G.\,Baulieu, 
 P.\,Calabria, 
 L.\,Caponetto, 
 C.\,Combaret, 
 R.\,Della\,Negra, 
 G.\,Grenier, 
 R.\,Han, 
 J-C.\,Ianigro, 
 R.\,Kieffer, 
 I.\,Laktineh, 
 N.\,Lumb, 
 H.\,Mathez, 
 L.\,Mirabito, 
 A.\,Petrukhin, 
 A.\,Steen, 
 W.\,Tromeur, 
 M.\,Vander\,Donckt, 
 Y.\,Zoccarato 
}\address{
Universit\'{e} de Lyon, Universit\'{e} Lyon 1, 
CNRS/IN2P3, IPNL, \\
4 rue E Fermi 69622,
Villeurbanne CEDEX, France
}

\author{%\centering 
E.\,Calvo~Alamillo, 
M.-C.\, Fouz, 
J.\,Puerta-Pelayo 
}\address{
CIEMAT, Centro de Investigaciones Energeticas, Medioambientales y Tecnologicas, Madrid, Spain 
}

\author{%\centering 
F.\,Corriveau
}\address{
Institute of Particle Physics of Canada and Department of Physics,\\
Montr\'{e}al, Quebec,
Canada H3A 2T8
}

\author{%\centering 
B.\,Bobchenko, 
M.\,Chadeeva, 
M.\,Danilov$^h$, %%\endnote{Also at MEPhI and at Moscow Institute of Physics and Technology}, 
A.\,Epifantsev, 
O.\,Markin, 
R.\,Mizuk$^h$, %%\endnotemark[9], 
E.\,Novikov, 
V.\,Popov, 
V.\,Rusinov, 
E.\,Tarkovsky
}\address{
Institute of Theoretical and Experimental Physics, \\ B. Cheremushkinskaya ul. 25,
RU-117218 Moscow, Russia
}

%%Nigel Watson 20131108 \author{%\centering 
%%Nigel Watson 20131108 N.\,Kirikova, 
%%Nigel Watson 20131108 V.\,Kozlov, 
%%Nigel Watson 20131108 P.\,Smirnov, 
%%Nigel Watson 20131108 Y.\,Soloviev
%%Nigel Watson 20131108  \\ \it
%%Nigel Watson 20131108 P.\,N.\, Lebedev Physical Institute,
%%Nigel Watson 20131108 Russian Academy of Sciences,
%%Nigel Watson 20131108 117924 GSP-1 Moscow, B-333, Russia
%%Nigel Watson 20131108 }

\author{%\centering 
D.\,Besson, P.\,Buzhan,
A.\,Ilyin, V.\,Kantserov, V.\,Kaplin, A.\,Karakash, E.\,Popova, V.\,Tikhomirov
}\address{
Moscow Physical Engineering Inst., MEPhI,
Dept. of Physics, \\
31, Kashirskoye shosse,
115409 Moscow, Russia
}

\author{%\centering 
C.\,Kiesling, 
K.\,Seidel, 
F.\,Simon, 
C.\,Soldner, 
L.\,Weuste 
}\address{
Max Planck Inst. f\"ur Physik,
F\"ohringer Ring 6,
D-80805 Munich, Germany
}

\author{%\centering 
M.\,S.\,Amjad, 
J.\,Bonis, 
S.\,Callier, 
S.\, Conforti\,di\,Lorenzo, 
P.\,Cornebise, 
Ph.\,Doublet, 
F.\,Dulucq, 
J.\,Fleury, 
T.\,Frisson, 
N.\,van der Kolk,  
H.\,Li$^i$, %%\endnote{Now at LPSC Grenoble}, 
G.\,Martin-Chassard, 
F.\,Richard, 
Ch.\,de la Taille, 
R.\,P\"oschl, 
L.\,Raux, 
J.\,Rou\"en\'e, 
N.\,Seguin-Moreau
}\address{
Laboratoire de l'Acc\'{e}l\'{e}rateur Lin\'{e}aire, Centre Scientifique d'Orsay,\\ 
Universit\'{e} de Paris-Sud XI, CNRS/IN2P3, \\
BP 34, B\^atiment 200, F-91898 Orsay CEDEX, France
}

\author{
 M.\,Anduze, 
 V.\,Balagura, 
 V.\,Boudry, 
 J-C.\,Brient, 
 R.\,Cornat, 
 M.\,Frotin, 
 F.\,Gastaldi,  
 E.\,Guliyev$^j$, 
 Y.\,Haddad, 
 F.\,Magniette, 
 G.\,Musat, 
 M.\,Ruan$^k$, 
 T.\,H.\,Tran, 
 H.\,Videau
}\address{
Laboratoire Leprince-Ringuet (LLR)  -- \'{E}cole Polytechnique, CNRS/IN2P3, \\F-91128 Palaiseau, France
}

%%Nigel Watson 20131108 \author{%\centering 
%%Nigel Watson 20131108 M.\,Anduze, 
%%Nigel Watson 20131108 V.\,Balagura, 
%%Nigel Watson 20131108 V.\,Boudry, 
%%Nigel Watson 20131108 J-C.\,Brient, 
%%Nigel Watson 20131108 R.\,Cornat, 
%%Nigel Watson 20131108 M.\,Frotin, 
%%Nigel Watson 20131108 F.\,Gastaldi,  
%%Nigel Watson 20131108 E.\,Guliyev, 
%%Nigel Watson 20131108 Y.\,Haddad, 
%%Nigel Watson 20131108 F.\,Magniette, 
%%Nigel Watson 20131108 G.\,Musat, 
%%Nigel Watson 20131108 M.\,Ruan, 
%%Nigel Watson 20131108 T.H.\,Tran, 
%%Nigel Watson 20131108 H.\,Videau
%%Nigel Watson 20131108  \\ \it
%%Nigel Watson 20131108  Laboratoire Leprince-Ringuet (LLR)  -- \'{E}cole Polytechnique, CNRS/IN2P3, F-91128 Palaiseau, France
%%Nigel Watson 20131108 }

\author{%\centering 
B.\,Bulanek, 
J.\,Zacek 
}\address{
Charles University, Institute of Particle \& Nuclear Physics, \\
V Holesovickach 2,
CZ-18000 Prague 8, Czech Republic  
}

\author{%\centering 
J.\,Cvach, 
P.\,Gallus, 
M.\,Havranek, 
M.\,Janata, 
J.\,Kvasnicka, 
D.\,Lednicky, 
M.\,Marcisovsky,  
I.\,Polak, 
J.\,Popule, 
L.\,Tomasek, 
M.\,Tomasek, 
P.\,Ruzicka, 
P.\,Sicho, 
J.\,Smolik, 
V.\,Vrba, 
J.\,Zalesak 
}\address{
Institute of Physics, Academy of Sciences of the Czech Republic, \\ 
Na Slovance 2,
CZ-18221 Prague 8, Czech Republic
}

\author{%\centering 
B.\,Belhorma, 
H.\,Ghazlane 
}\address{
Centre National de l'Energie, des Sciences et des Techniques Nucl\'{e}aires, \\
B.P. 1382, R.P. 10001, Rabat, Morocco
}

\author{%\centering              
K.\,Kotera, 
H.\,Ono$^l$,
T.\,Takeshita, 
S.\,Uozumi$^\spadesuit$
}\address{
Shinshu Univ.\,,
Dept. of Physics,
3-1-1 Asahi,
Matsumoto-shi, Nagano 390-8621,
Japan
}

\author{D.\,Jeans$^\spadesuit$}
%\author{D.\,Jeans\corref{cor}}
%\ead{jeans@icepp.s.u-tokyo.ac.jp}
%\cortext[cor]{Corresponding author}
\address{
Department of Physics, Graduate School of Science, The University of
Tokyo, \\ 7-3-1 Hongo, Bunkyo-ku, Tokyo 113-0033, Japan
}

\author{
S.\,Chang, A.\,Khan, D.\,H.\,Kim, D.\,J.\,Kong, Y.\,D.\,Oh
}\address{
Department of Physics, Kyungpook National University, Daegu, 702-701,
Republic of Korea
}

\author{
M.\, G\"otze, 
J.\,Sauer, 
S.\,Weber, 
C.\,Zeitnitz
}\address{
Bergische Universit\"{a}t Wuppertal,
Fachbereich C Physik,\\
Gaussstrasse 20,
D-42097 Wuppertal, Germany
}

% $^\spadesuit$ Corresponding author\newline
% E-mail: \email{jeans@icepp.s.u-tokyo.ac.jp}
% }

\author{
\flushleft
\footnotesize
\makebox[8pt][l]{$^\spadesuit$}Corresponding authors. E-mail: {\tt jeans@icepp.s.u-tokyo.ac.jp, satoru@knu.ac.kr}\\
\makebox[8pt][l]{$^a$}Now at Northern Illinois University\\
\makebox[8pt][l]{$^b$}Also at University of Texas, Arlington\\
\makebox[8pt][l]{$^c$}Deceased\\
\makebox[8pt][l]{$^d$}On leave from ITEP\\
%\makebox[8pt][l]{$^d$}Now at DESY\\
\makebox[8pt][l]{$^e$}Also at DESY\\
\makebox[8pt][l]{$^f$}Also at Argonne National Laboratory\\
\makebox[8pt][l]{$^g$}Now at University of Sussex, Physics and Astronomy Department,\\
\makebox[8pt][l]{}Brighton, Sussex, BN1 9QH, UK\\
\makebox[8pt][l]{$^h$}Also at MEPhI and at Moscow Institute of Physics and Technology\\
\makebox[8pt][l]{$^i$}Now at LPSC Grenoble\\
\makebox[8pt][l]{$^j$}Now at TRIUMF, Vancouver, BC, Canada\\
\makebox[8pt][l]{$^k$}Now at IHEP, Beijing, China\\
\makebox[8pt][l]{$^l$}Now at Nippon Dental University, Niigata, Japan
}

%\theendnotes

%NIM

\end{frontmatter} %NIM

\section{Introduction}

A future high energy lepton collider~\cite{ILC_TDR_summary, CLIC_CDR}, 
running up to TeV-scale energies,
will play a crucial role in unravelling the nature of
% the precision measurement of
the Higgs sector and 
potential physics beyond the Standard Model (SM), as well
as providing high precision measurements of SM processes involving
W, Z bosons, and the top quark.
Since these latter particles will decay largely into multi-jet final
states, precise jet energy measurement is a critical issue.
A jet energy resolution of 3\%
% added 25/02
over a wide range of jet energies, 
significantly better than
what has been achieved in past and present collider detectors, will allow
hadronic decays of W and Z bosons to be effectively distinguished.
Event reconstruction by Particle Flow Algorithms (PFA)~\cite{PFA-brient, PandoraPFA}
has the potential to achieve this level of jet energy resolution.
A number of detector designs, optimised for the use of PFA, 
are being developed~\cite{ILC_TDR_detectors, CLIC_CDR-physDet}. 
These designs require calorimeters of very fine granularity,
which allow the identification of single particles inside jets within the calorimeter,
an essential requirement for PFA.
The electromagnetic (EM) calorimeter (ECAL) 
for such an approach 
requires a lateral granularity
of $5\times5$ -- $10\times10~\mathrm{mm}^2$~\cite{ILD_LOI}, typically giving $10^7$--$10^8$ readout channels. 
The single particle energy resolution of the calorimeters is less critical in the PFA approach: an ECAL energy resolution
of $15\%/\sqrt{E(\mathrm{GeV})}$ gives a rather small contribution to the jet energy resolution~\cite{PandoraPFA}.
% DANIEL added 24/12/2013
The large number of channels in such a detector requires a design with rather simple and robust construction methods
to enable the detector to be built within reasonable time and manpower resources.

The CALICE scintillator strip-based ECAL (\mbox{ScECAL}) achieves the required granularity with a scintillator strip structure.
Each strip is individually read out by a Multi Pixel Photon Counter
(MPPC, a silicon photon detector produced by Hamamatsu Photonics K.K.~\cite{MPPC}).
Although plastic scintillators have been widely used in calorimeters,
this is the first time that a highly granular calorimeter has been made using scintillator strips.
Such an ECAL has a smaller cost than
alternative technologies using silicon sensors (e.g.~\cite{SiW-commission}).
% , in particular a
% lower production cost. 
% and a smaller number of readout channels.
The MPPC has promising properties for the \mbox{ScECAL}: a small size (active area of $1 \times 1~\mathrm{mm}^2$ in a package of 
% \mbox{$3 \times 2 \times 1.3~\mathrm{mm}^3$}), 
\mbox{$4.2 \times 3.2 \times 1.3~\mathrm{mm}^3$}), 
excellent photon counting ability, 
low cost and low operation voltage ($\sim$80~V), 
% DANIEL added 24/12/2013
with disadvantages of temperature-dependent gain, saturation at high
light levels, and a relatively high dark noise rate.
The use of tungsten absorber material minimises the Moli\`ere radius of the calorimeter, an important 
aspect for the effective separation of particle showers required by PFA reconstruction.
% DANIEL added 24/12/2013
The chosen strip geometry allows a reduction in the number of readout channels, while maintaining
an effective granularity given by the strip width, by the use of appropriate reconstruction
algorithms. One such algorithm, know as the Strip Splitting Algorithm~\cite{SSA}, has been developed and
demonstrated to perform well in jets expected at ILC.

A first \mbox{ScECAL} prototype consisting of 468 channels was constructed and then tested
in February and March 2007 using positron beams provided by the DESY-II electron synchrotron~\cite{DESY_testbeam}.
The aim of this experiment was to demonstrate the
feasibility of a scintillator strip ECAL with MPPC readout 
for a future detector, with sufficiently good energy resolution for PFA-based jet energy reconstruction.
Large-scale tests of Hamamatsu MPPCs in a real detector
had not yet been performed before the present tests.

% 
% A first \mbox{ScECAL} prototype consisting of 468 channels was constructed and then tested
% in February and March 2007 using positron beams provided by the DESY-II electron synchrotron~\cite{DESY_testbeam}.
% The main aim of this prototype was to gain experience in the construction, operation and calibration of an ECAL 
% using scintillator strip and MPPC technologies. 
% Large-scale tests of Hamamatsu MPPCs in a real detector had not yet been performed before the present tests.
% This operation of this prototype highlighted several aspects of the design requiring further development for future prototypes.
% The exposure to positron beams allowed the detector to be calibrated, and first measurements of its energy response to be made.
% The DESY positron test beam lines are very convenient for detector commissioning, they however have certain limitations
% for calorimetric tests, in particular the limited energy range and somewhat poorly defined beam momentum spread.
% These factors, together with the prototype's small size and a number of identified design limitations,
% limit the accuracy with which this first prototype's measured performance can be extrapolated to a complete ScECAL system.
% Such a full characterisation of ScECAL technology requires the development of a second, larger prototype,
% implementing the improvements prompted by the experience of operating this first prototype. This larger prototype
% will be used to better test the technology's energy response, its pattern recognition capabilities,
% and to make comparisons with the predictions of detector simulations.

In this paper we present an analysis of the
energy resolution and linearity of the first \mbox{ScECAL} prototype, using data collected with 1--6~GeV/c positron beams.
In Section~\ref{sec:ScECALprototype}, 
the \mbox{ScECAL} prototype is described. 
Section~\ref{sec:MPPCsaturation} presents the measurement and correction of the non-linearity of the MPPC response, and 
in Section~\ref{sec:DESY_BT} we describe the instrumentation on the DESY beam line and present the data analysis and its results. 
A Monte Carlo simulation of the prototype is presented in Section~\ref{sec:simulation}.
The results are discussed and summarised in Section~\ref{sec:summary}.

\section{ScECAL prototype}\label{sec:ScECALprototype}

\begin{figure}[t]
\center{
\includegraphics[height=0.36\textwidth]{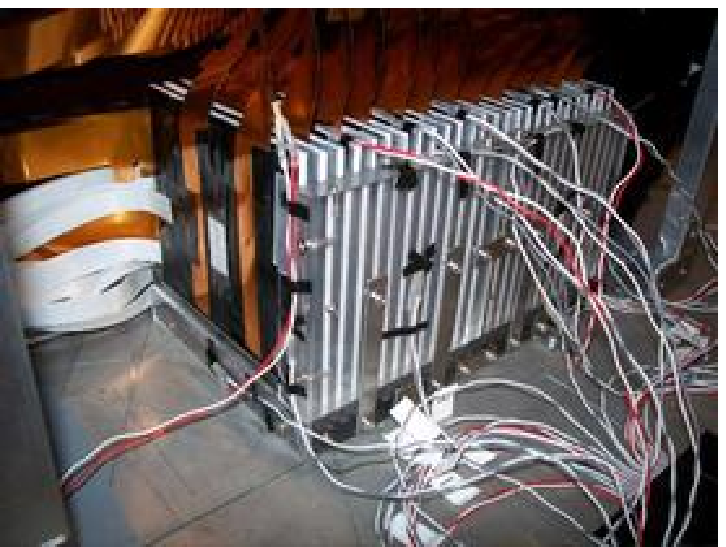}
\hspace{0.04\textwidth}
\includegraphics[height=0.36\textwidth]{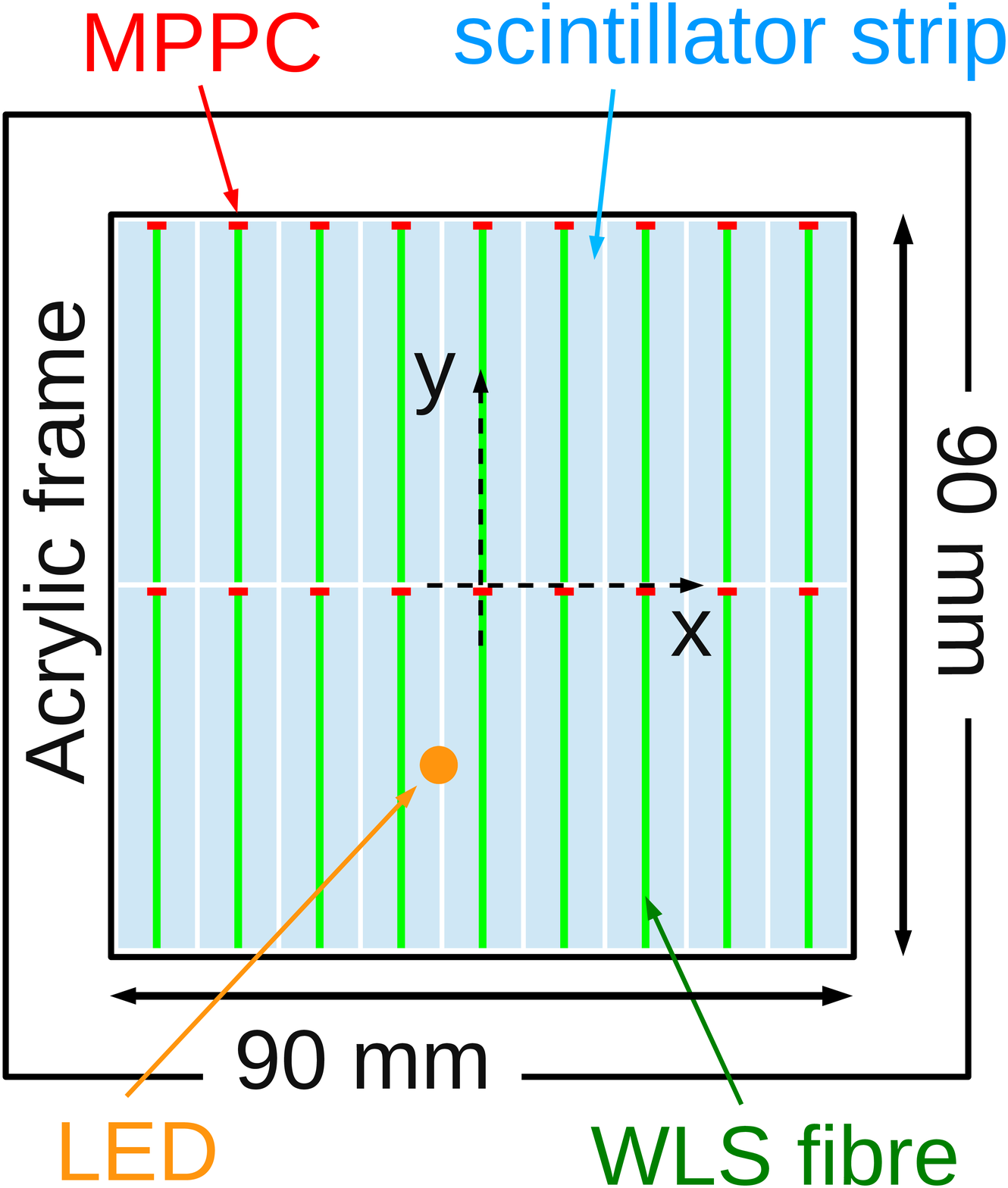}
}
\caption{
\textbf{Left:} Photograph of the \mbox{ScECAL} prototype. The 26 active layers are seen in the clear acrylic support structure.
The golden-coloured flat cables are MPPC readout cables and the twisted pair cables in the foreground
are connected to the temperature sensors. The white flat cables connect the LEDs of the calibration system.
\textbf{Right:} Structure of a type-F detector layer, showing the two mega-strips, each divided into nine strips,
the positions of the WLSFs, MPPCs, and the calibration LEDs. The definition of the coordinate system is also shown.
}
\label{fig:ECALmodule}
\end{figure}

The \mbox{ScECAL} prototype is shown in Fig.~\ref{fig:ECALmodule}.
It consisted of 26 pairs of 3~mm thick scintillator and 3.5~mm thick absorber layers, 
placed in an acrylic support structure.
The absorber material was composed of 82\% tungsten, 13\% cobalt and about 5\% carbon, 
% added daniel 17/01/2014
with an estimated radiation length of 5.3~mm.
% end daniel
The effective Moli\`ere radius of the prototype was 22~mm.

\begin{figure}[t]
\center{
\includegraphics[height=0.2\textwidth]{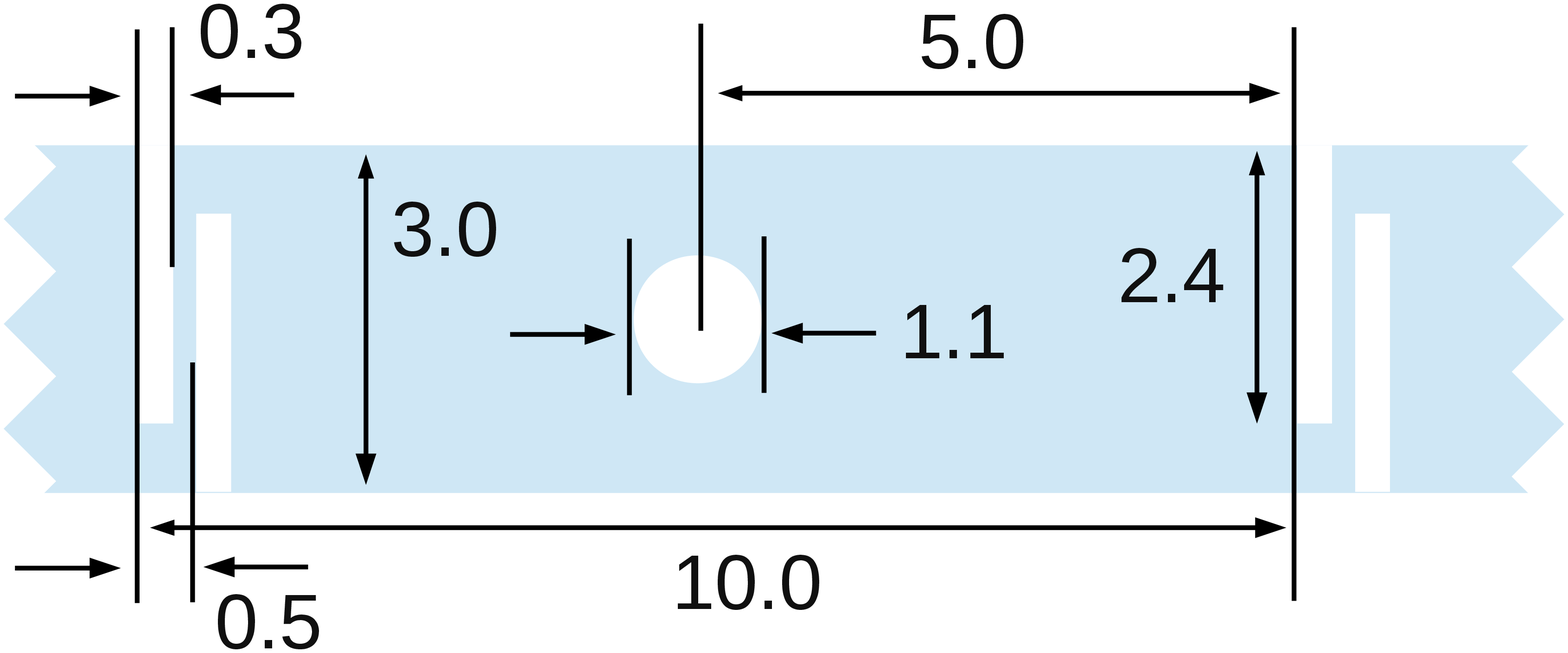}
\hspace{0.1\textwidth}
\includegraphics[height=0.2\textwidth]{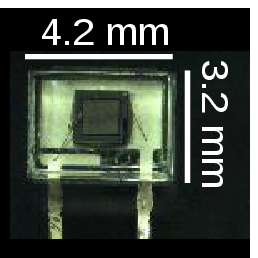}
}
\caption{
\textbf{Left:} Cross-section of one strip of the type-F mega-strip structure (all dimensions are in mm).
The design of type-D mega-strips is the same, except without the hole.
\textbf{Right:} Photograph of a MPPC. 
The package size is 
% $3 \times 2 \times 1.3~\mathrm{mm}^3$, 
$4.2 \times 3.2 \times 1.3~\mathrm{mm}^3$, 
and the 1600 pixels are contained in an active area of $1 \times 1~\mathrm{mm}^2$.
}
\label{fig:mega-strip}
\end{figure}

Each scintillator layer consisted of two $45\times90~\mathrm{mm}^2$ ``mega-strip'' structures consisting of nine $45\times10~\mathrm{mm}^2$ strips.
% DANIEL added 24/12/2013
Compared to a design based on individual strips, a mega-strip design allows simpler alignment and construction techniques to be used when 
producing a large-scale detector with tens of millions of channels; on the other hand it has the disadvantage of optical cross-talk between 
strips, which complicates the interpretation of collected data.
The mega-strips were produced by machining holes and grooves in a 3~mm-thick
Kuraray SCSN38 plastic scintillator plate, as shown in Fig.~\ref{fig:mega-strip}. 
White polyethylene terephthalate (PET) film (thickness $\sim 130~\mu$m)
was 
% inserted 
% DANIEL added 24/12/2013
placed (not glued) 
into the grooves to optically isolate
adjacent strips. Test bench studies measured an optical cross-talk between strips of around 10\%. 
Within each layer, two mega-strips were placed side-by-side separated by a strip of the same PET film.
%Both sides of each pair of mega-strips were
% The mega-strips were 
%covered by a sheet of 3M radiant mirror reflector film to increase the amount of light collected by the MPPCs.
The two sides of each layer were covered first by a 
sheet of 3M radiant mirror reflector film to increase the amount of light collected by the MPPCs, and then a
black vinyl sheet to block external light.
Two types of detection layers were produced: ``type-F(ibre)'' with a 1~mm diameter 
Kuraray Y-11 wavelength shifting optical fibre (WLSF) running along the length of the strip,
and ``type-D(irect)'' without the WLSF or its associated hole.
% DANIEL added 24/12/2013
The WLSF was held in place by its natural curvature within the straight hole, without the use of any glue.
The presence of the WLSF improves the response uniformity 
along the strip length (from a non-uniformity of around 30\% for type-D to 15\% for type-F), but
increases the complexity 
% (and therefore also cost) % daniel would leave this out...
of scintillator strip 
manufacture and assembly.

Each mega-strip was read out by nine, 1600~pixel MPPCs soldered onto a custom-made flat readout cable
and mounted in holes at the end of each strip.
% DANIEL added 24/12/2013
The alignment of the strips to the MPPCs was controlled at the level of $\pm 100 \mu m$. 
To simplify the construction procedure, no special optical
coupling (glue or grease) was used between the fibre and MPPC.
The distribution of the MPPCs' break-down voltage had a mean of around 74~V and a variation (RMS) of around 0.5\%.
% Measurements made before the test beam showed that 
The variation of the MPPC pixel capacitance was around 5\%.
A uniform operating over-voltage (difference between the operation and break-down voltage) 
could therefore be used within each module without introducing large gain fluctuations\footnote{
The MPPC pixel gain $G$ is related to the pixel capacitance $C$ and over-voltage $\Delta V$ by 
$G = C \cdot \Delta V / e = q / e$, where $e$ is the electron charge and $q$ the single pixel output charge~\cite{MPPC}.}.
% DANIEL moved to later 26/12/2013
%An over-voltage of 2.9~V (3.2~V) was applied to MPPCs in type-F (type-D) layers.
%The higher over-voltage used in type-D results in an improved MPPC light detection efficiency,
%partially compensating for the lower scintillation light collection efficiency of this strip type.

Two 13-layer modules were constructed, each using a single mega-strip type.
These two modules were tested in two configurations: in the ``\FD\ configuration'',
the type-F module was placed directly upstream of the type-D module, while in
the ``\DF\ configuration'', their order was reversed.
Since most of the EM shower energy (88--75\% for 1--6~GeV/c electrons)
is contained in the first half of the \mbox{ScECAL} prototype, the characteristics of each
configuration were dominated by the upstream module.
Scintillator layers were placed in two alternating orientations, 
with horizontally and vertically aligned strips, giving an effective granularity of $10 \times 10~\mathrm{mm^2}$.
The total active volume was about $90\times90\times200$~mm$^2$, 
with a thickness of around 
% daniel change 17/01/14
% 22
17 
% end daniel
radiation lengths, and was read out by 468 MPPCs.
After assembly, five channels were found not to provide signals,
probably due to problems with signal line connections.
Since these correspond to only around 1\% of all channels, and were located 
mostly at the edge of the detector,
their effect on the overall calorimeter performance is expected to be negligible.

Signals from the MPPCs were read by the front-end (FE) electronics developed for the 
CALICE analogue hadron calorimeter (AHCAL) prototype~\cite{AHCAL_commissioning},
consisting of the pre-amplifier, shaper and sample-and-hold circuit within the ILC-SiPM ASIC~\cite{ILC-SIPM},
followed by a 16 bit ADC in the CALICE readout cards.
The digitised data from the ADCs were transmitted to and further processed by the 
CALICE data acquisition system.
The FE also provided an adjustable bias voltage to the MPPCs.

%treated by the CALICE front end electronics (CALICE FE) \cite{calice-fee-Dauncey}\cite{calice-fee}\cite{SiW-commission}
%the very front end (VFE) shaper and amplifier ASIC chips, and digitised by the CALICE readout cards (CRC).
%, allows channel-by-channel 
%bias voltage tuning which allows uniform gain to be acheived for all MPPCs. % not really used, so don't mention?
Two gain modes of the CALICE FE were used: a high-gain mode (with a gain of $\sim 90$~mV/pC) giving sufficient sensitivity to measure single photo-electron signals
but a small dynamic range (linear up to around 11.5 pC), and a low-gain mode ($\sim 8$~mV/pC) with lower sensitivity but a sufficiently large dynamic range (linear up to 190 pC)
to measure large calorimeter signals~\cite{beni-thesis}.
%This FE provided two gain modes for signal readout:
%the high-gain mode was sufficiently sensitive to measure a single photo-electron signal, 
%however the dynamic range was not sufficient to measure the large signals expected within EM showers, while
%the low-gain mode had a lower sensitivity but a larger dynamic range in order to measure these large signals.
All beam data were collected in the low-gain mode, 
%The low-gain mode was used to collect all beam data, 
while the high-gain mode was used to make calibration measurements using MPPC single pixel signals, 
as described in Section~\ref{sec:MPPCsaturation}.

% DANIEL added 26/12/2013
The applied MPPC over-voltage was chosen to give a large enough gain to clearly resolve
single photon peaks when using the high gain readout mode.
While satisfying this requirement, essential for the gain calibration, as low as possible
an over-voltage is preferred to maximise the dynamic range when running in low-gain mode with electron showers,
and to minimise the MPPC inter-pixel cross-talk and dark noise rate. Neither of these undesirable
MPPC features had a significant adverse effect on this prototype's performance.
% DANIEL moved from earlier 26/12/2013
An over-voltage of 2.9~V (3.2~V) was applied to MPPCs in type-F (type-D) layers.
The higher over-voltage used in type-D results in an improved MPPC light detection efficiency,
partially compensating for the lower scintillation light collection efficiency of this strip type.

The detector was equipped with an LED calibration system used to perform in-situ measurements of the MPPC gain.
Each layer was equipped with one blue SMD LED (package size $\rm 1.6 \times 0.8 \times 0.4 ~mm^3$, from the SEIWA electric company),
connected via a 0.3~mm-thick flexible flat cable. 
The LED was placed just off the centre of one of the two mega-strips, over the boundary between two strips.
The light was transmitted to the scintillator by a small hole in the black vinyl sheet and reflector film, and 
propagated to adjacent strips due to the non-perfect optical isolation between strips.
Around 25\% of strips could be calibrated using this system.

\section{MPPC saturation correction}
\label{sec:MPPCsaturation}

The MPPC response is intrinsically non-linear due to its finite
number of pixels, which leads to a saturation of its response at high light levels. 
If an input light pulse 
is shorter than the MPPC recovery time (measured to be $\sim$4~ns for the MPPCs used in this experiment),
the MPPC response can be parameterised by
\begin{equation}
N_{\mathrm{fired}}
(N_{\mathrm{p.e.}}) = N_{\mathrm{pix}} (1-e^{-N_{\mathrm{p.e.}}/N_{\mathrm{pix}}}),
\label{eq:MPPCsaturation}
\end{equation}
where $N_{\mathrm{fired}}$ denotes the number of fired pixels, $N_{\mathrm{pix}}$ the total number of MPPC pixels, and 
$N_{\mathrm{p.e.}}$ the number of photo-electrons created (the product of the number of incoming photons and the MPPC photon detection efficiency).
This function is approximately linear for $N_{\mathrm{p.e.}} \ll N_{\mathrm{pix}}$, 
however for larger signals the MPPC output begins to saturate, with a maximum response of $N_{\mathrm{fired}} = N_{\mathrm{pix}}$ as the number
of photo-electrons approaches infinity.
The MPPCs used in this experiment have $N_{\mathrm{pix}} = 1600$.

If the input light pulse is longer than the typical MPPC recovery time,
% (measured to be $\sim$4~ns for the MPPCs used in this experiment), 
the effective dynamic range is
increased due to the possibility of a single pixel firing several times within the same light pulse. 
This effect occurs particularly in type-F strips due to the relatively slow decay time of the 
WLSF (measured to be $\sim$8~ns). 
The decay time of the scintillator itself was measured to be $\sim$2~ns, significantly shorter than the MPPC recovery time,
so type-D strips are not expected to show a strong enhancement of the dynamic range.
To take this enhancement into account, we replace $N_{\mathrm{pix}}$ by an effective number of pixels 
$N_{\mathrm{pix}}^{\mathrm{eff}}$ in the MPPC response function (\ref{eq:MPPCsaturation}), and measure this effective pixel number
in both strip types.

In the following %subsections 
we discuss the procedure used
to correct for this saturation effect: measurement of the single pixel signal,
the MPPC response curve, and the formulation of the saturation correction.

\subsection{Single pixel signal}\label{sec:dvalue}

\begin{figure}[t]
\center{
\includegraphics[width=0.45\textwidth]{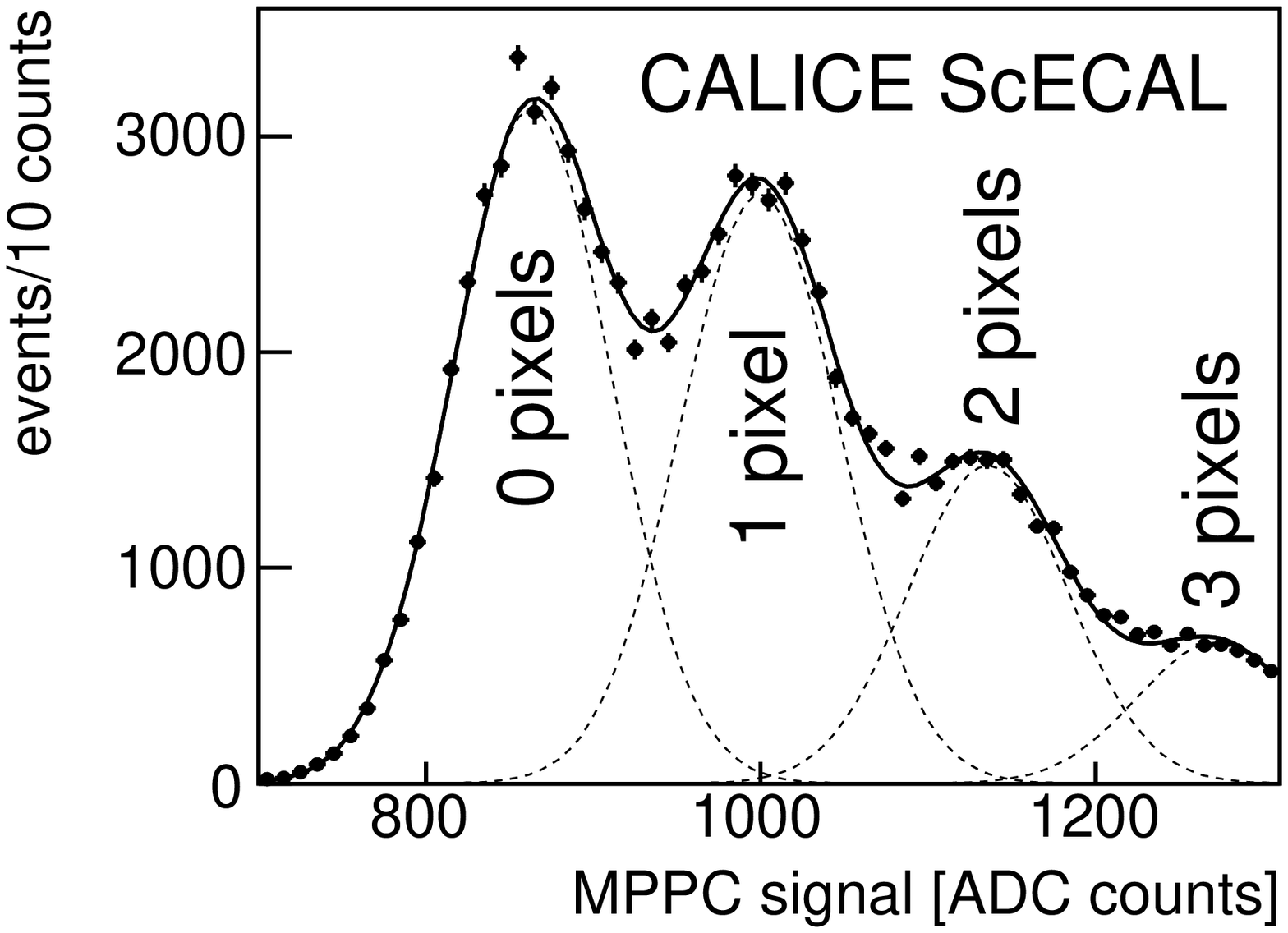}
\includegraphics[width=0.45\textwidth]{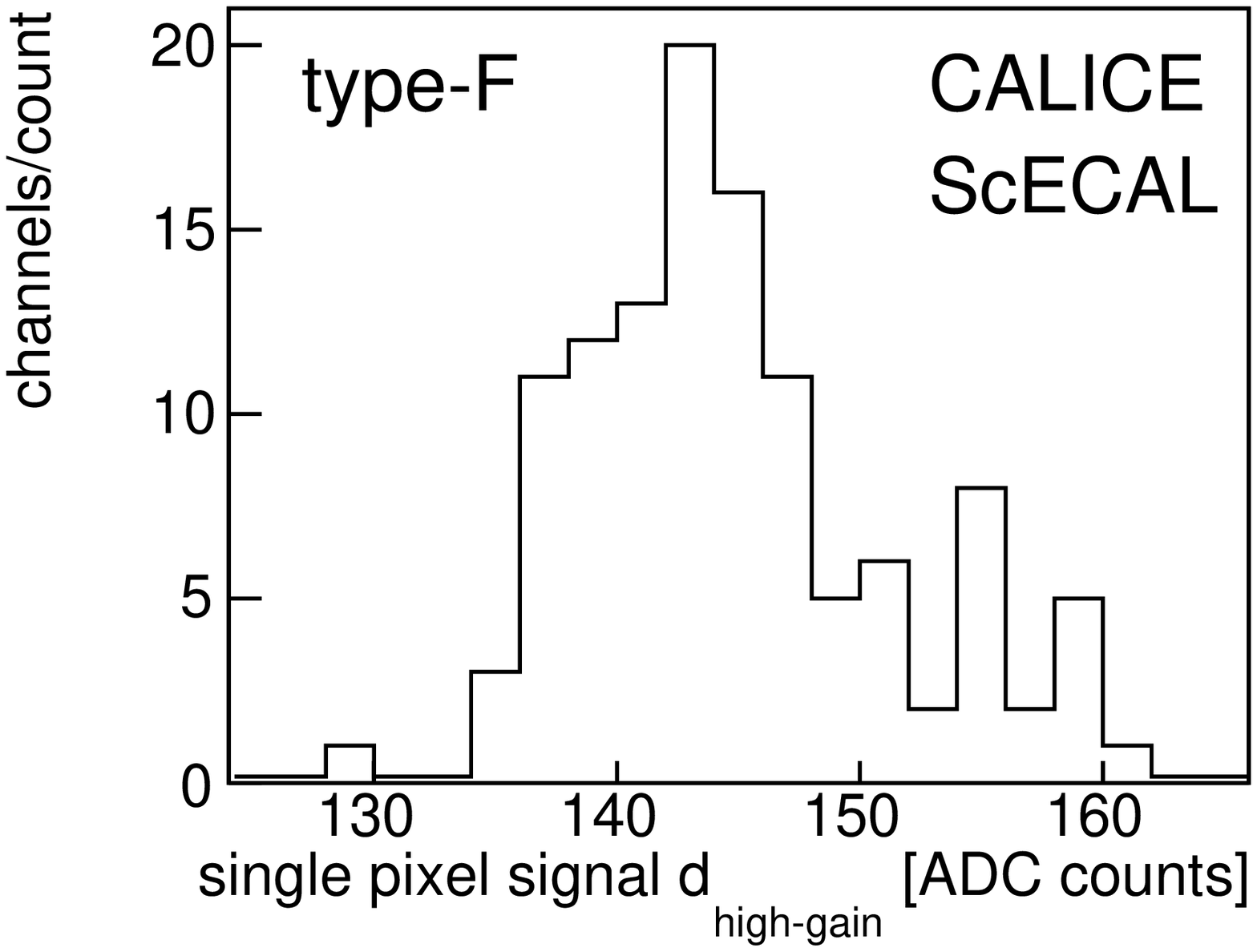} \\
}
\caption{
\textbf{Left:} A typical MPPC output spectrum taken with the LED system, showing the 0-, 1-, 2- and 
% 3-photo-electron 
3-fired-pixel
peaks and the results of the fit.
\textbf{Right:} Distribution of the measured single pixel signals $d_{\mathrm{high-gain}}$ in the type-F module 
with an MPPC over-voltage of 2.9~V.
}
\label{fig:dval}
\end{figure}

The signal $d$ produced by a single fired pixel is related to the 
% pixel capacitance $C$ and over-voltage $\Delta V$ by 
% $d = k \cdot C \cdot \Delta V$, daniel 25/02 for ref#2
MPPC gain $G$ by
$d = k \cdot q = k \cdot G \cdot e$, 
where $k$ is the conversion factor from signal charge to ADC counts.
It was measured (in terms of ADC counts) in the 25\% of strips accessible to the LED calibration system.
Using a low power LED signal, and with the FE electronics in high-gain mode, characteristic signal distributions were observed,
consisting of a pedestal and a few 
% photo-electron 
fired-pixel
peaks. An example is shown in Fig.~\ref{fig:dval} (left).
These spectra were fitted by a function of the form
\begin{equation}
 f(x) = \sum_{i=0}^{3} A_i\ e^{-\frac{(x-i\cdot d - \mu_0)^2}{2\sigma^2}},\nonumber
\end{equation}
in which the Gaussian with index $i$ corresponds to the $i$-th 
% photo-electron
fired-pixel
peak. 
The parameter % dj added 25/02
$\mu_0$ is the central value of the first, zero 
% photo-electron 
fired pixel
peak, which corresponds to the pedestal.
The distance between 
% photo-electron 
pixel
peaks, $d$, corresponds to the signal due to a single fired pixel. It is taken to be constant,
assuming linear MPPC response in this low-signal region.
$A_i$ and $\sigma$ are, respectively, the normalisations and width of the Gaussian peaks.
The width of each peak is dominated by electronics noise rather than variations in pixel gain,
so a common width was assumed for all peaks. 
$A_i$, $d$, $\mu_0$ and $\sigma$ were treated as free parameters in the fit.
% \\ \textbf{DANIEL: what about $\mu_0$???. I write that it is also free parameter.} \\
The result of such a fit is also shown in Fig.~\ref{fig:dval}~(left). %, as is the distribution of the measured single pixel signals.
%
%Due to time pressures during the test beam, (\textbf{and maybe to problems with led system?}) the single pixel signal was measured 
%for only $\sim$25\% of the channels.
%\\ \textbf{DANIEL: 25\% of all channels, or 25\% of LED-equipped channels? or is it the same? I cannot remember....Is this 25\% for successful tests, or all attempts?}\\

Figure~\ref{fig:dval}~(right) shows the distribution of the single pixel signals measured in the 
$\sim25\%$ of accessible channels in the type-F module using an over-voltage $\Delta V$ of 2.9~V.
A second measurement of a similar number of MPPCs was made at an over-voltage of 3.7~V.
The single pixel signal at $\Delta V = 3.2$~V, as used in the \mbox{type-D} module, was taken to be an
interpolation between these two measurements, assuming a linear dependence of the single pixel signal on the over-voltage.
The averages and variations (RMS) of the single pixel signal distributions at the two chosen over-voltages were determined to be
%and 
%Using the same over-voltage as during data taking, the average and root mean square (RMS) of the single pixel signals in the 
%two modules were measured to be
\begin{eqnarray}
d^{\mathrm{type-F}}_{\mathrm{high-gain}} & = & 144.9 \pm 6.4\ \mbox{(RMS) ADC counts at $\Delta V^{\mathrm{type-F}} = 2.9$~V} \nonumber \\
d^{\mathrm{type-D}}_{\mathrm{high-gain}} & = & 151.6 \pm 8.3\ \mbox{(RMS) ADC counts at $\Delta V^{\mathrm{type-D}} = 3.2$~V. \nonumber}
\end{eqnarray}

% daniel moved to previous section
% Before the test beam we had confirmed that variation of pixel capacitance
% in the MPPCs used to build the prototype was reasonably small, less than 4~\%.
% We could therefore set a uniform over-voltage within a module, while still acheiving
% a relatively constant gain.
% %In a same type of module, over-voltage is uniformly set to achieve same gain.
% A common over-voltage of 2.9~V (3.2~V) was used for all MPPCs in module type-F (type-D).
% A higher over-voltage was used in type-D to compensate for the lower scintillator
% light output by increasing the MPPC detection efficiency.

\begin{figure}[t]
\center{
\includegraphics[width=0.45\textwidth]{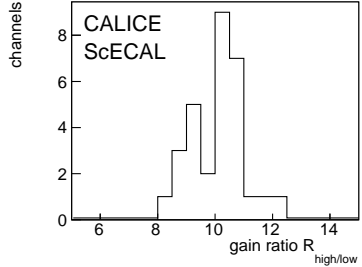}
}
\caption{
Distribution of the gain ratio $\mathrm{R}_{\mathrm{high/low}}$ in the 30 measured channels.
}
\label{fig:intercalib}
\end{figure}

To apply the single pixel signals measured in high gain mode to the
% saturation correction of 
test beam data collected in low gain mode, the single pixel signal
must be translated from high to low gain mode using the ratio of the two gains.
This gain ratio $\mathrm{R_{high/low}}$ was measured by comparing the MPPC signals produced by a medium-strength LED signal of around 150 photo-electrons
using both high and low gain modes. % The ratio of the two measurements gives the gain ratio $\mathrm{R_{high/low}}$.
This ratio was measured in 30 channels, shown in Fig.~\ref{fig:intercalib}, yielding a mean value and RMS of
\begin{equation}
\mathrm{R_{high/low} = \langle ADC_{high-gain} \rangle / \langle ADC_{low-gain} } \rangle = 10.08 \pm 0.95~(\mathrm{RMS}). \nonumber
\end{equation}
This average gain ratio was applied to both module types to calculate the
low gain single pixel signals shown in Table~\ref{tab:dvalues_lowgain_Npix}.
The uncertainty on $d_{\mathrm{low-gain}}$ takes into account the
variation of the measurements of $d_{\mathrm{high-gain}}$ and of the gain ratio.

\begin{table}[t]

\begin{center} 
\begin{tabular}{ccc}\hline
Module type & $d_{\mathrm{low-gain}}$  & $N_{\mathrm{pix}}^{\mathrm{eff}}$ \\
            &  [ADC counts] & \\
\hline
%type-F & 14.4 $\pm$ 1.5 & 2072.8 \\
%type-D & 15.8 $\pm$ 1.7 & 1677.2 \\
type-F & 14.4 $\pm$ 1.5 & 2073 \\
type-D & 15.8 $\pm$ 1.7 & 1677 \\
\hline
\end{tabular}
\end{center}
\caption{The average low-gain single pixel signal $d_{\mathrm{low-gain}}$ and the effective number of pixels $N_{\mathrm{pix}}^{\mathrm{eff}}$ in the two module types.}
\label{tab:dvalues_lowgain_Npix}
\end{table}

\subsection{MPPC response curve}

\begin{figure}[t]
\center{
\includegraphics[height=0.35\textwidth]{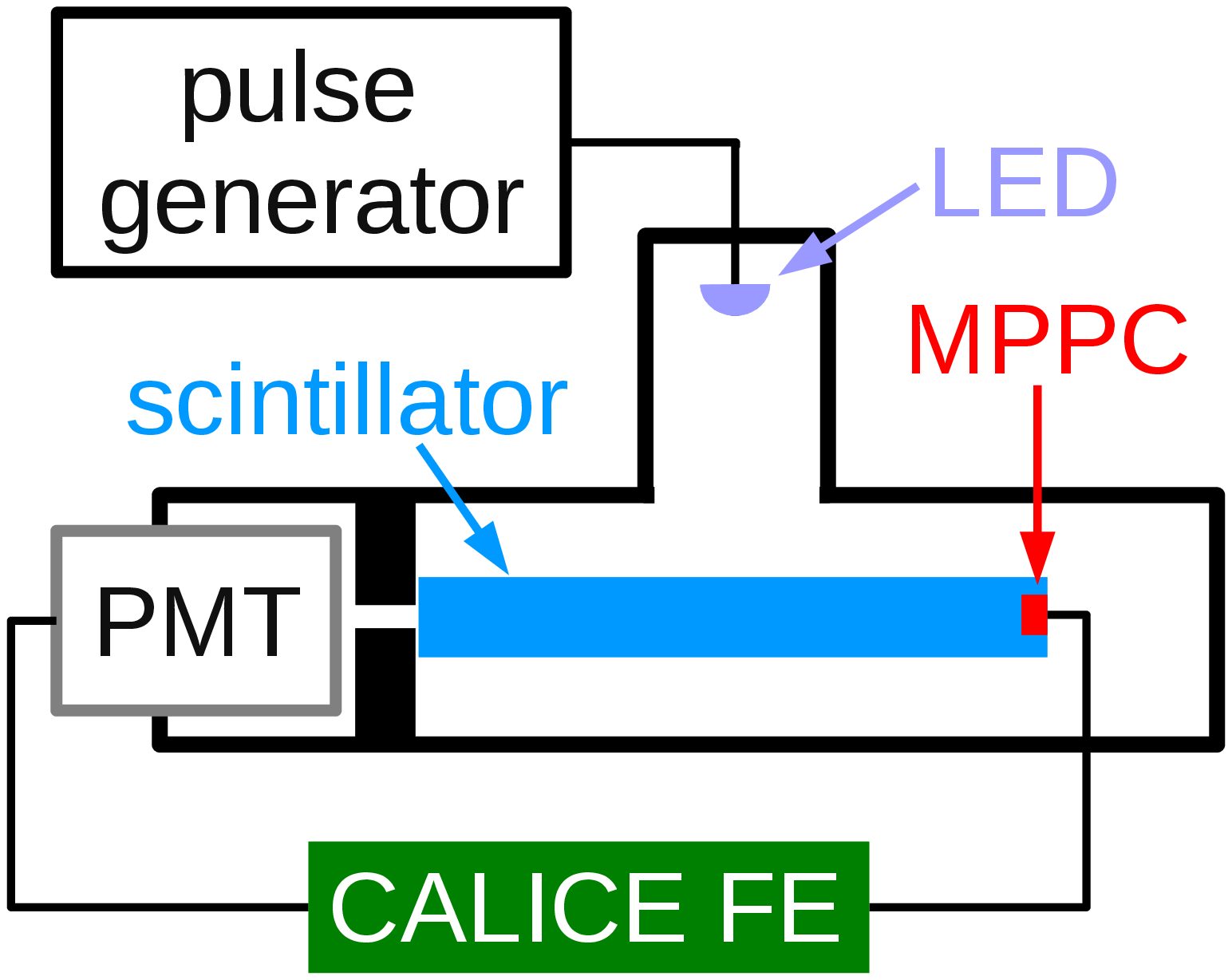}
\includegraphics[height=0.35\textwidth]{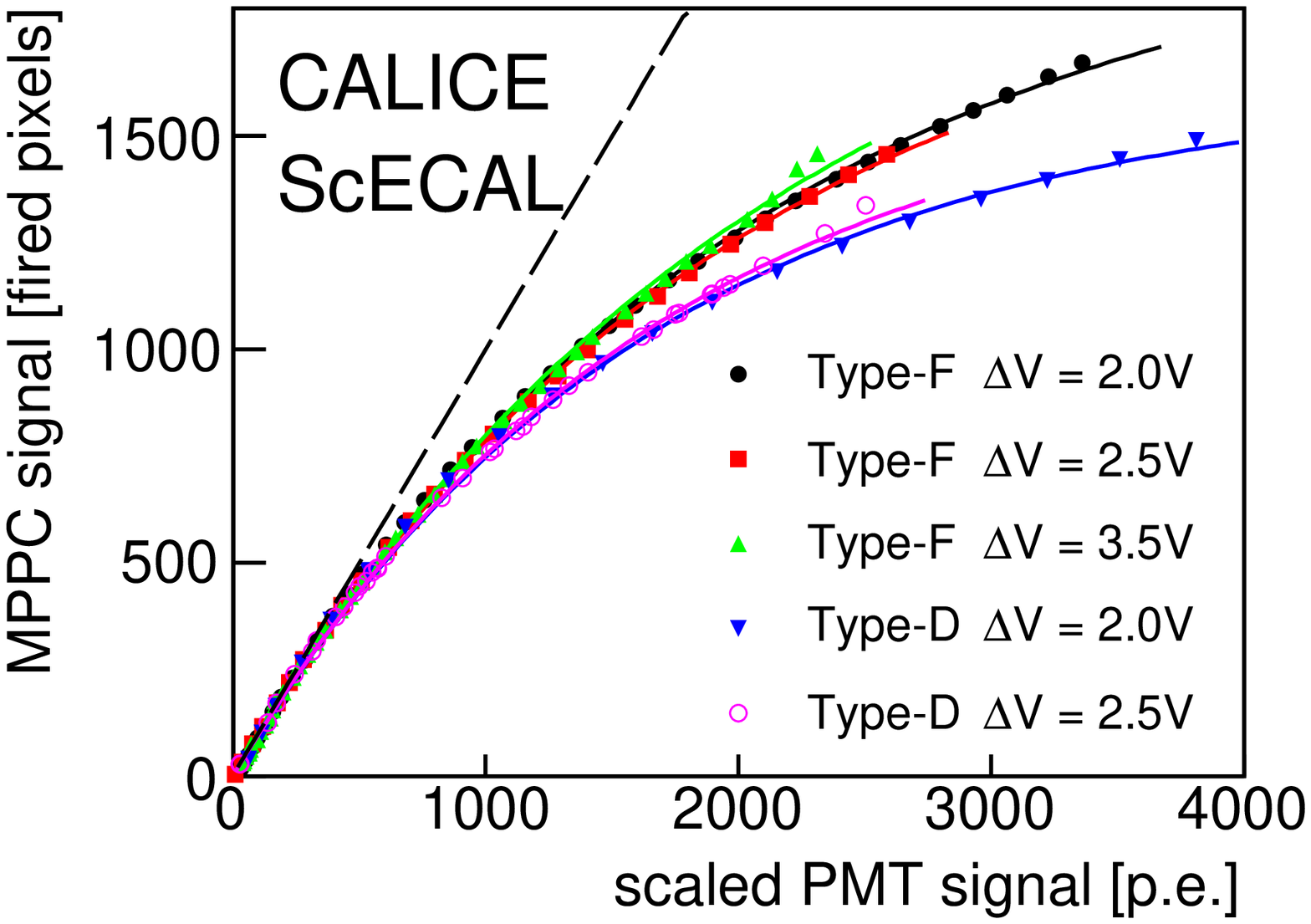}
%/home/jeans/Documents/CALICE_scecal_paper/CALICEnotes/draft/saturation_curve_final.eps}
}
\caption{
\textbf{Left:} The MPPC response curve measurement setup.
\textbf{Right:} Measured MPPC response curves when using the two types of scintillator strips at different MPPC over-voltages $\Delta V$.
The dashed line shows a linear response. Typical effective pixel numbers for the two types are given in Table~\ref{tab:dvalues_lowgain_Npix}.}
\label{fig:responsecurves}
\end{figure}

The response of the 
% scintillator strip and 1600~pixel MPPC system 
scintillator strip -- MPPC system 
was measured using the dedicated apparatus shown in Fig.~\ref{fig:responsecurves}.
An ultraviolet LED was used to inject a 
% 10~ns light pulse 
light pulse with a length of a few~ns
into the centre of a scintillator strip. 
The generated scintillation light was measured at both ends of the strip, at one end by 
a photomultiplier tube (PMT) from Hamamatsu Photonics and at the other by 
a MPPC of the same type as used in the prototype.
The scintillator strip had dimensions of $45\times10\times3~\mathrm{mm}^3$, and was made of the
same material as the mega-strips used in the prototype. 
Two types of strips were tested, with and without a WLSF.
The signals from the PMT and MPPC were read out using the CALICE FE electronics in low-gain mode.
The signal from the PMT, which does not suffer from saturation at high light levels, 
is proportional to the light produced in the scintillator, and therefore also to the light input to the MPPC.
The comparison of the PMT and MPPC responses at different light intensities
allows the extraction of the MPPC response curve.

The MPPC signal (in ADC counts) was converted to the number of fired pixels $N_{\mathrm{fired}}$ using 
the single pixel signal $d_{\mathrm{low-gain}}$ obtained as described in Section~\ref{sec:dvalue}:
\begin{equation}
 N_{\mathrm{fired}} = \mathrm{ADC}_{\mathrm{MPPC}}/ d_{\mathrm{low-gain}}.\nonumber
\end{equation}
The number of effective pixels $N_{\mathrm{pix}}^{\mathrm{eff}}$ was then obtained by measuring the
dependence of the number of fired MPPC pixels on the PMT signal.
To extract the number of pixels $N_{\mathrm{pix}}^{\mathrm{eff}}$, 
this dependence was fitted by a function of the form
\begin{equation}
N_{\mathrm{fired}} = N_{\mathrm{pix}}^{\mathrm{eff}}
\left(1-e^{-p\cdot\mathrm{ADC}_{\mathrm{PMT}}/N_{\mathrm{pix}}^{\mathrm{eff}}}\right), \nonumber
%\label{eqn:fitfunc}
\end{equation}
where $\mathrm{ADC}_{\mathrm{PMT}}$ is the photomultiplier
response (in ADC counts). The multiplicative factor $p$,
% daniel 12/2/2014
calculated separately for each measured response curve,
is used to convert the PMT signal in ADC counts
to the number of 
% MPPC photo-electrons, 
% fired MPPC pixels,
photo-electrons,
compensating for the different gains and efficiencies of the two devices.

Figure~\ref{fig:responsecurves} shows the response curves of the two scintillator types at different bias voltages.
The type-F strip shows a larger enhancement of $N_{\mathrm{pix}}^{\mathrm{eff}}$ than
the type-D strip, due to the slow decay time of the WLSF emission.
No strong dependence on the applied bias voltage was
observed with either scintillator type. This is expected, since only the time structure
of the input light determines the enhancement of the MPPC dynamic range.
The average of $N_{\mathrm{pix}}^{\mathrm{eff}}$ values measured at different
bias voltages was used in the saturation correction. 
These averages for the two strip types are shown in Table~\ref{tab:dvalues_lowgain_Npix}.
The statistical uncertainty on $N_{\mathrm{pix}}^{\mathrm{eff}}$ estimated by the fit procedure
is 
% smaller than
at most % dj, jp 8/5/14
a few pixels in all cases.
The fact that only a rather small enhancement is observed in the type-D strip demonstrates that
the length of the input light pulse is somewhat shorter than the MPPC recovery time, and that the
present results can be applied to test beam data, in which the energy deposition is 
essentially instantaneous.

\subsection{Saturation correction}\label{sec:corrfunc}

The effects of MPPC saturation in beam data were corrected using
the results outlined above. The signal from each MPPC, $\mathrm{ADC_{raw}}$
was converted to the corresponding number of fired pixels by using the
appropriate single pixel signal ($d$): $N_{\mathrm{fired}} = {\mathrm{ADC}_\mathrm{raw}}/{d_{\mathrm{low-gain}}}$.
The saturation-corrected number of photo-electrons $N_{\mathrm{p.e.}}$ was estimated as
\begin{equation}
 N_{\mathrm{p.e.}} = -N_{\mathrm{pix}}^{\mathrm{eff}}
 \ln\left(1-\frac{N_{\mathrm{fired}}}{N_{\mathrm{pix}}^{\mathrm{eff}}}\right),\nonumber
\end{equation}
which was then converted back to a corrected ADC signal 
\begin{equation}
\mathrm{ADC}_{\mathrm{corrected}} = N_{\mathrm{p.e.}} \cdot d_{\mathrm{low-gain}}.\nonumber
\end{equation}

\subsection*{Estimation of systematic effects}
There are two possible sources of uncertainty in the application of the 
saturation correction, the imperfect knowledge of $d_\mathrm{low-gain}$ and $N_\mathrm{pix}^\mathrm{eff}$. 
Effects due to $d_\mathrm{low-gain}$ depend only on the MPPC and FE electronics, and
are therefore expected to have similar behaviour in different detector configurations and regions.
A single uncertainty was therefore used for these effects. 
Measurements of $d_\mathrm{high-gain}$ were made in around 25\% of all channels, and
the gain ratio was measured in thirty ($\sim 6\%$) channels. The means of these two sets of measurements
were used to calculate an average value for $d_\mathrm{low-gain}$ in the two strip types.
The uncertainty on these average values (reflected in Table~\ref{tab:dvalues_lowgain_Npix}) is dominated by the $\sim$10\% variation in the 
measurements of $\mathrm{R_{high/low}}$.
The uncertainty was estimated by comparing the results obtained when varying
$d_\mathrm{low-gain}$ above and below its nominal value by its uncertainty 
with the results obtained when using the nominal value.

Effects due to $N_\mathrm{pix}^\mathrm{eff}$ are expected to depend also on the strip type, 
since the two types produce a different number of photo-electrons per MIP, 
giving rise to different saturation characteristics as a function of MIPs.
Different uncertainties were therefore used for the two detector configurations.
The nominal value of $N_\mathrm{pix}^\mathrm{eff}$ was taken from the measurement of only a 
single strip of each type. The relative uncertainty on the fitted $N_\mathrm{pix}^\mathrm{eff}$ is small $\lsim 10^{-3}$.
Although the MPPCs used in this experiment were measured to have rather similar properties, 
some residual variations can be expected, due to non-uniform properties of the 
MPPC, scintillator strip, WLSF, and the couplings between them. The scale of the effect of these variations 
on $N_\mathrm{pix}^\mathrm{eff}$ is not very precisely known.
The length of the light pulse used to measure $N_\mathrm{pix}^\mathrm{eff}$ is also not very well known,
which can affect the precision on the effective pixel number.
To estimate the uncertainty due to these effects, 
a comparison was made of the 
%the dj/jp 8/5/14
detector performances estimated when assuming different
values of $N_\mathrm{pix}^\mathrm{eff}$, uniformly applied to all detector channels.
$N_\mathrm{pix}^\mathrm{eff} = 1600, 1677$, and 2073 were considered, corresponding respectively to the physical
number of pixels, and the measured effective pixel numbers for type-D and type-F strips.
Half of the larger difference seen when comparing $N_\mathrm{pix}^\mathrm{eff} = 1600$ and 2073, and
$N_\mathrm{pix}^\mathrm{eff} = 1600$ and 1677, 
was taken as a conservative estimate of the systematic uncertainty 
due to imprecise estimation of the effective pixel number.

%To estimate the uncertainty due to this, the stochastic and constant terms ($r_{s,c}$)
%estimated when applying uniform values for $N_\mathrm{pix}^\mathrm{eff}$ of 1600, 1677, and 2073 were considered.
%Half of the larger of $| r_{s,c} (1600) - r_{s,c}(2073) |$ and $| r_{s,c}(1600) - r_{s,c}(1677) |$ 
%was taken as a conservative
%estimate of the systematic uncertainty due to this effect.
%on the stochastic and constant terms of the two configurations.

%The difference in detector performance observed when using $N_\mathrm{pix}^\mathrm{eff}=1600$ and
%$N_\mathrm{pix}^\mathrm{eff}=2073$
%The size of the uncertainty was estimated by comparing the results obtained 
%when applying the saturation correction with three different values of $N_\mathrm{pix}^\mathrm{eff}$ to all channels:
%1600 (the physical number of pixels), 2073 (as measured in the type-F strip), and
%1677 (as measured in the type-D strip).
%assuming
%the measured $N_\mathrm{pix}^\mathrm{eff}$ and $N_\mathrm{pix}^\mathrm{eff}=1600$. Half of the
%difference between these two cases was taken as a conservative estimate of the uncertainty.

\section{Test beam at DESY}\label{sec:DESY_BT}

\begin{figure}[t]
\center{
\includegraphics[width=\textwidth]{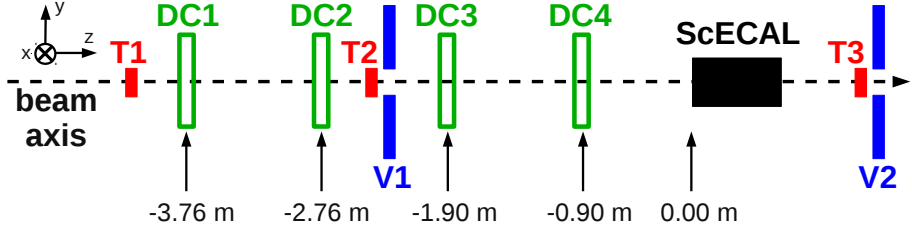}
}
\caption{
Sketch of the beam line instrumentation, showing the layout of the 
four pairs of drift chambers (DC1-4), three trigger counters (T1-3) and the two veto counters (V1-2) relative to the 
\mbox{ScECAL} prototype. (Not to scale.)
%\textbf{DANIEL needs the scintillator positions.}
}
\label{fig:beamline}
\end{figure}

This test beam experiment was performed at the DESY-II electron synchrotron (at DESY, Hamburg, Germany)
using positron beams with momenta of 1, 2, 3, 4, 5, and 6~GeV/c.
A sketch of the beam line instrumentation is shown in Fig.~\ref{fig:beamline}.
The trigger (T1-3) and veto (V1, 2) counters are scintillator detectors of size $3\times3$~cm$^2$ %daniel removed thickness
and $20\times20$~cm$^2$, respectively. The veto counters have a $2\times2$~cm$^2$
hole at their centre. These counters were each read out by two photomultipliers (PMT).
Two trigger and one veto counter were placed upstream of the prototype, and one trigger and
one veto counter were placed downstream of the \mbox{ScECAL}.
The beam line was also instrumented with four pairs of drift chambers (DC1-4). 
Each DC has an active area of $\sim72 \times 72~\mathrm{mm}^2$
and measures either the $x$ or $y$ coordinate.
Each pair of DCs measures both the $x$ and $y$ position of beam particles.
The \mbox{ScECAL} was placed on a movable stage, allowing it to be moved with 0.1~mm precision in the plane
normal to the beam.
The readout electronics were housed in a crate placed next to the prototype.
The temperature was monitored by a number of temperature sensors
placed on and around the detector.

During beam data taking, the acquisition was triggered by a coincidence of signals from one PMT of trigger counter T1 and one PMT of T2.
The pedestal distribution of all \mbox{ScECAL} channels was monitored regularly during data taking by recording 1000 randomly 
triggered events every 10000 beam events.
In each readout channel, the mean signal recorded in these randomly triggered events 
was subtracted from the following set of beam trigger events.

\subsection{Calibration}\label{sec:MIPcalib}

The absorber plates were removed from the detector to collect data used for calibration. 
Calibration runs were performed several times during the test beam period, using
a 3~GeV/c positron beam. The detector was moved during calibration 
runs to ensure that all strips were exposed to a sufficient number of particles.
An event pre-selection required that the signals from all trigger and veto counters, both up- and down-stream of the \mbox{ScECAL}, 
were consistent with a single, non-showering particle passing through the \mbox{ScECAL}.

\begin{figure}[t]
\center{
\includegraphics[width=0.6\textwidth]{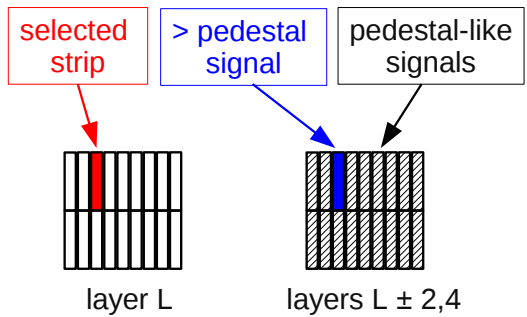}
}
\caption{Illustration of the ECAL-based MIP selection. 
For a given strip in a particular layer L (highlighted on the left), the
strips on surrounding layers of the same orientation (right) were required to be consistent (hatched)
or inconsistent (filled) with a pedestal signal, depending on the strip position.
}
\label{fig:MIP_sel}
\end{figure}

Each strip was then individually calibrated
with data for which the drift chamber track reconstruction showed that a particle passed through the strip.
Events satisfying this requirement are in the following referred to as DC-selected.
An additional event selection was applied to these events, using the data in other ECAL layers, 
as illustrated in Fig.~\ref{fig:MIP_sel}.
The signals on the two preceding and two succeeding \mbox{ScECAL}
layers with the same orientation as the target strip layer were considered. If such a layer was not available, 
the requirement was ignored.
% in particular for layers at the beginning or end of the calorimeter, only available layers satisfying these requirements were used. 
Within these layers, the signals on
the ``same-position'' strips (the strips in the same $x, y$ position as the target strip) were
required to 
% be inconsistent with a pedestal-only signal, 
measure a signal at least 80~ADC counts above pedestal,
and the strips in other positions on these same layers were 
required to have a signal 
% consistent with a pedestal-only signal.
no larger than 80 ADC counts above pedestal. Events satisfying these criteria are referred to as ECAL-selected.

\begin{figure}[t]
\center{
\includegraphics[width=0.45\textwidth]{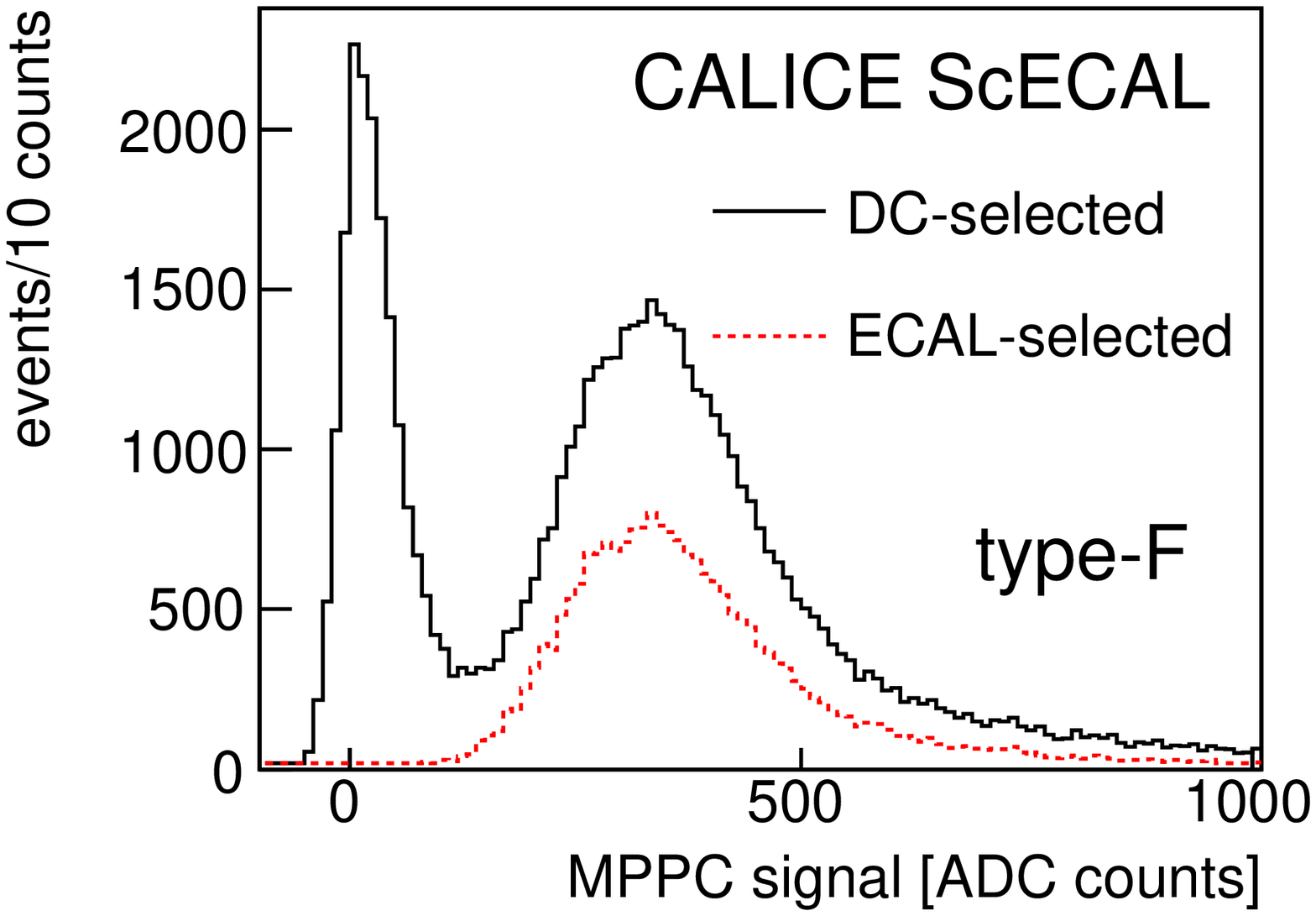}
\includegraphics[width=0.45\textwidth]{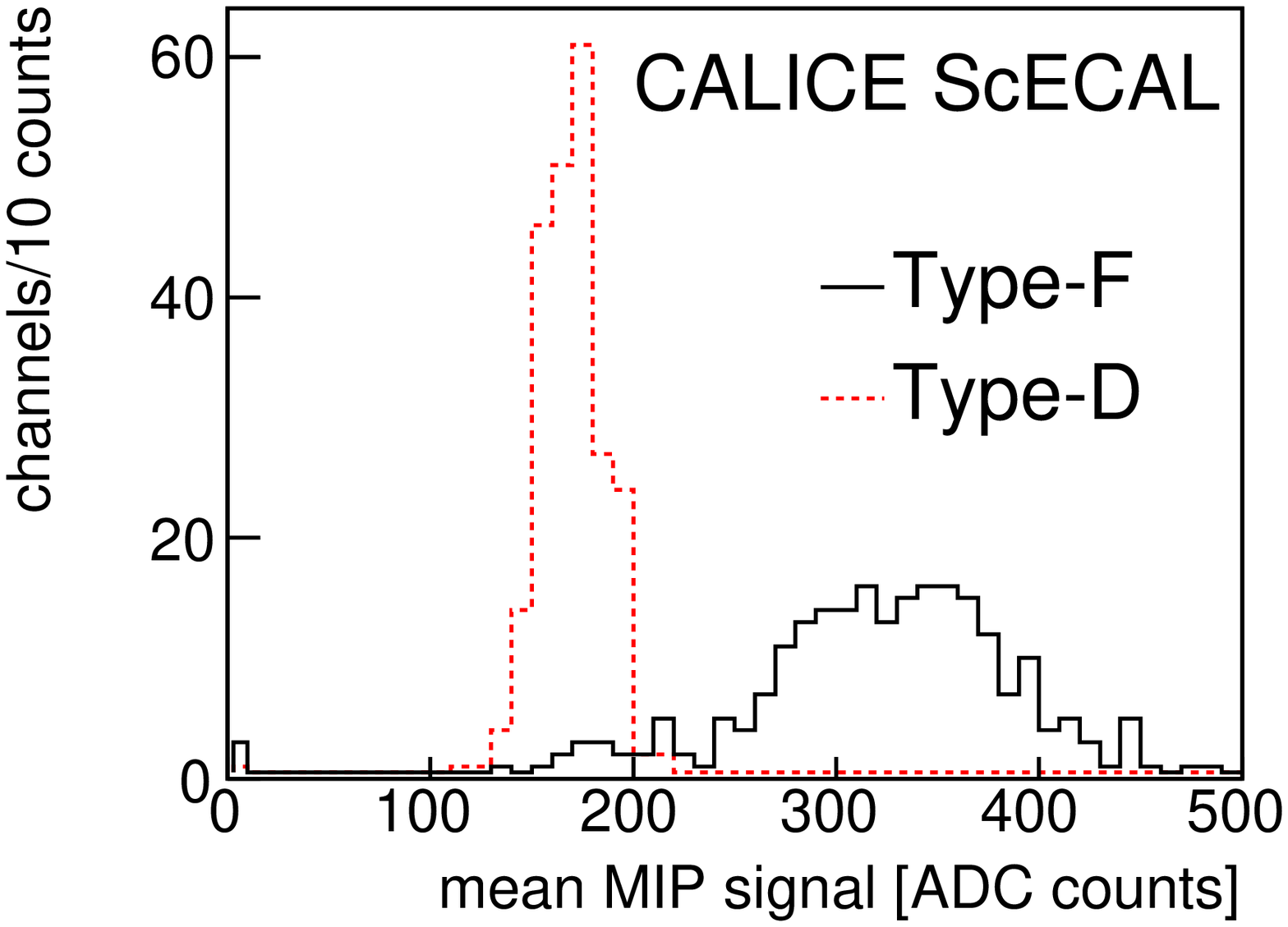}
}
\caption{
\textbf{Left:} ADC distribution for MIP runs, after drift chamber (DC) and ECAL based selections, 
for a representative 
type-F % dj/jr 8/5/14
strip.
% with WLSF readout.
\textbf{Right:} Distributions of the measured calibration constants for the two strip types.}
\label{fig:MIP_calib}
\end{figure}

The signal distributions of a representative 
%WLSF 
type-F % dj/jr 8/5/14
channel after these two event selections are shown in
Fig.~\ref{fig:MIP_calib}.
The mean values %\footnote{The mean was used rather than the most probable value of a Landau function fit for reasons of robustness.} 
of the final ECAL-selected distributions were used as the
channel-by-channel calibration constants translating the signal in ADC counts
to Minimum Ionizing Particle equivalent units (MIPs)\footnote{We loosely use ``MIP'' to denote the most probable signal 
produced by a 3~GeV/c positron when crossing one layer of scintillator. 
The signal induced by a true Minimum Ionising Particle is slightly different.}.
The mean MIP signals measured in all strips are plotted in Fig.~\ref{fig:MIP_calib}.
Two distributions are seen, due to the two scintillator types. 
Type-F strips give, on average, a larger signal than type-D
(this depends on differences in light collection efficiency and MPPC gain between the two types),
and also a significantly larger dispersion (23\% for type-F, compared to 11\% for type-D). 
The dispersion for type-F is consistent with our expectation of effects
% DANIEL added 24/12/2013
due to the $\mathcal{O}(100 \mu m)$ variations
% due to variations 
in the relative positioning of the WLSF and MPPC, in particular variations in the 
distance between the end of the WLSF and the MPPC package.

\subsubsection*{Estimation of systematic effects}
A pseudo-experiment method was used to estimate the uncertainties arising from the statistical
uncertainty of the measured MIP calibration constants.
In each pseudo-experiment the calibration constant of each channel was randomly varied according to a 
Gaussian distribution whose standard deviation was the measured statistical uncertainty of 
the channel's calibration. The variation seen within an ensemble of one hundred pseudo-experiments
was taken as the systematic uncertainty due to this effect.

\subsection{Temperature dependence}\label{sec:tempCorr}

\begin{figure}[t]
\center{
\includegraphics[width=0.45\textwidth]{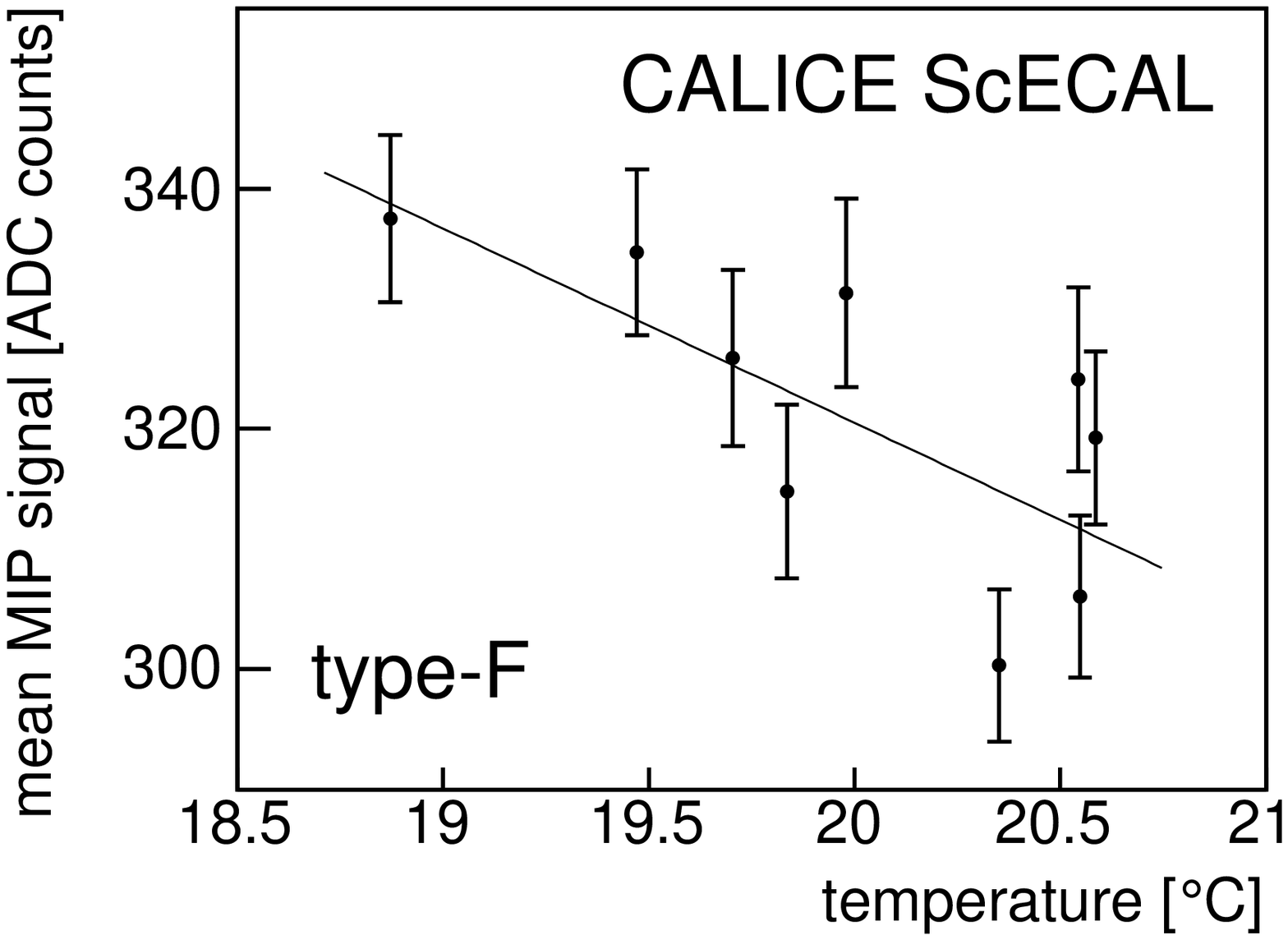}
\includegraphics[width=0.45\textwidth]{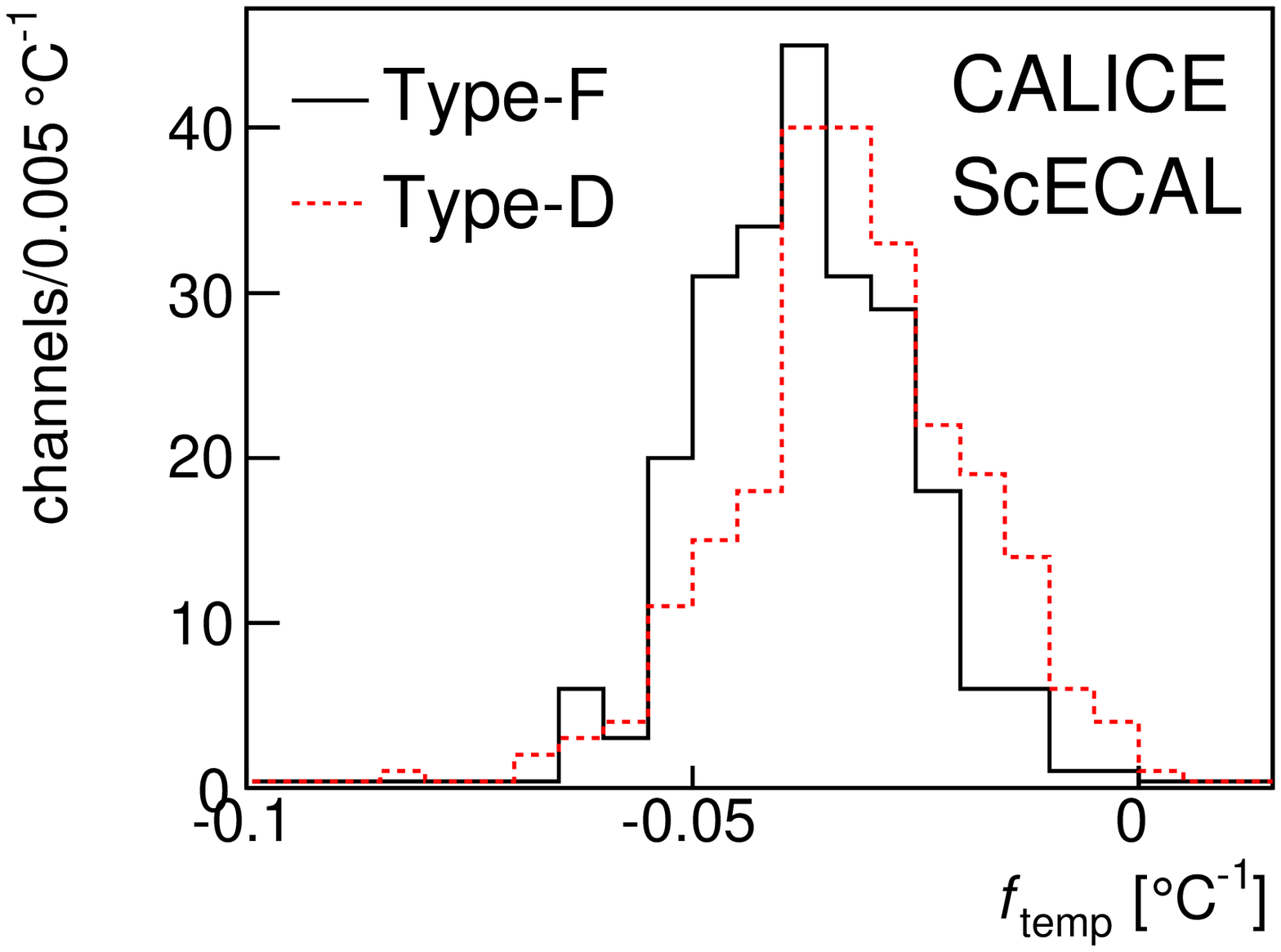}
}
\caption{
\textbf{Left:} Variation of the MIP response with temperature in a typical channel.
\textbf{Right:} Distribution of the temperature coefficients $f_{\mathrm{temp}}$ for all channels.}
\label{fig:calib_temp_coeff}
\end{figure}

Since the MPPC gain $G$ depends strongly on the temperature $T$ \linebreak
(${\delta G/\delta T \sim -2\%/^\circ\mathrm{C}}$ at $20^\circ\mathrm{C}$~\cite{MPPC}), correction of this
effect is essential.
Calibration runs were taken several times
during the test beam period, 
%with a fairly representative spread of temperature.
covering the temperature range encountered during EM shower data taking.
The mean MIP response was measured individually in each calibration run,
and used to extract its dependence on the temperature.
Figure~\ref{fig:calib_temp_coeff} shows this variation for a representative channel.
The dependence of each channel's response on temperature $c(T)$ was fitted with a linear function
\begin{equation}
c(T)  = c(T_0) \cdot ( 1 + f_{\mathrm{temp}} \cdot (T - T_0) )\nonumber
\end{equation}
where the reference temperature $T_0$ was chosen to be $20^\circ$C.
The distributions of the fitted temperature coefficients,
$f_{\mathrm{temp}}$, for the two strip types
are shown in Fig.~\ref{fig:calib_temp_coeff}.
The MIP response decreases with increasing temperature
in a similar way for type-F and type-D strips, as expected.
In the analysis of EM shower events described in
Section~\ref{sec:reslin}, the response of each strip was calibrated using
a temperature-dependent calibration function determined by these fitted functions.

\begin{figure}[t]
\center{
\includegraphics[width=0.6\textwidth]{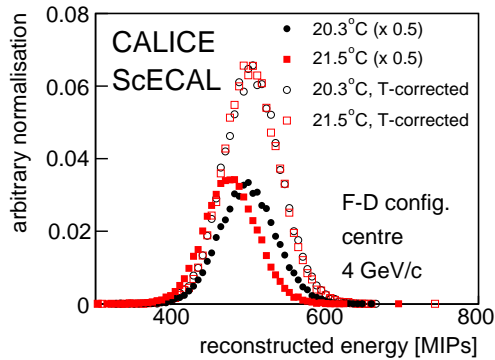}
}
\caption{
The measured energy distributions of EM shower data in the central region of the \FD\ detector configuration
in two runs at a beam momentum of 4~GeV/c. 
The closed (open) symbols show the response before (after) the application 
of the temperature correction. The uncorrected curves have been scaled by 50\% to aid visibility.
(The cross-talk correction has not been applied.)
}
\label{fig:tempcorrdemo}
\end{figure}

To demonstrate the efficacy of this correction, in Fig.~\ref{fig:tempcorrdemo} we show the 
reconstructed energy distributions of central EM shower 
events collected by the \FD\ configuration detector
at a beam momentum of 4~GeV/c. The data were collected during two data-taking periods, each of around
20 minutes, separated by around 14 hours. The mean temperatures of the prototype during the two runs were 20.3 and 
$21.5^\circ$~C. 
The application of the temperature correction reduces the relative difference between the mean energy response
in the two data-taking periods from $(4.51\pm0.06)\%$ to $(0.17\pm0.06)\%$.
%
% split into 2 figures (for ref#2)
%
% \begin{figure}[t]
% \center{
% \includegraphics[width=0.45\textwidth]{dj_tempcorrdemo2.eps}
% \includegraphics[width=0.45\textwidth]{dj_xtalkcorrdemo2.eps}
% }
% \caption{
% \textbf{Left:} 
% The measured energy distributions of EM shower data in the central region of the \FD detector configuration
% in two runs at a beam momentum of 4~GeV/c. 
% The closed (open) symbols show the response before (after) the application 
% of the temperature correction. The uncorrected curves have been scaled by 50\% to aid visibility.
% (The cross-talk correction has not been applied.)
% \textbf{Right:} 
% The reconstructed energy after the temperature correction (closed circles) and after an additional cross-talk (XT) correction (open squares), 
% for the same sample of central events collected at 4~GeV/c by the \FD detector configuration.
% }
% \label{fig:tempcorrdemo}
% \end{figure}
% 

\subsubsection*{Estimation of systematic effects}
A pseudo-experiment method was used to estimate the uncertainties arising from the statistical
uncertainty of the measured temperature dependence.
In each pseudo-experiment the temperature correction factor $f_{\mathrm{temp}}$ of each channel was 
randomly varied according to a Gaussian distribution whose standard deviation was the 
measured statistical uncertainty of $f_{\mathrm{temp}}$.
The variation seen within an ensemble of one hundred such pseudo-experiments
was taken as the systematic uncertainty due to this effect.

\subsection{Optical cross-talk}\label{sec:xtalk}

\begin{figure}[t]
\center{
\includegraphics[width=0.45\textwidth]{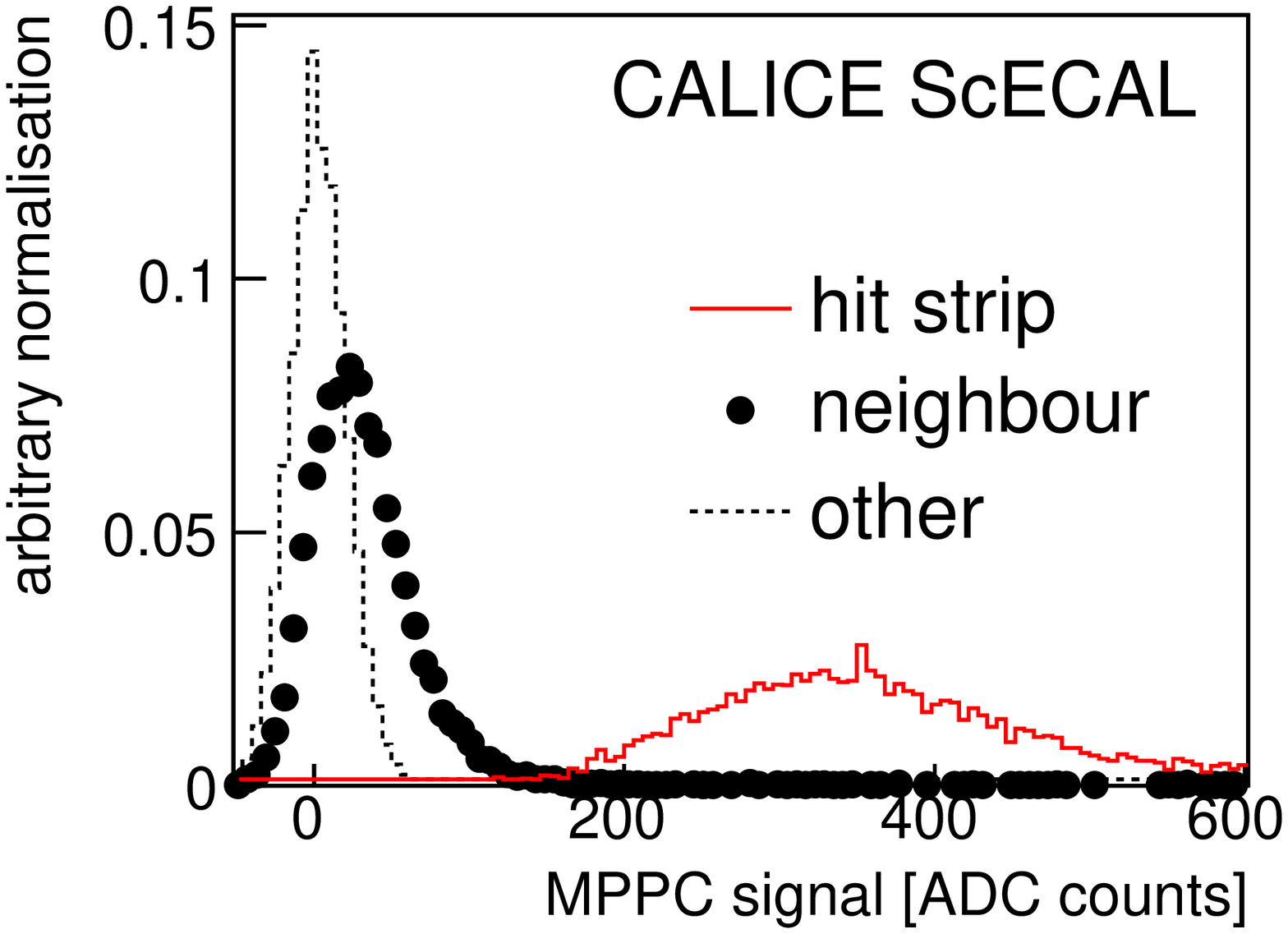}
\includegraphics[width=0.45\textwidth]{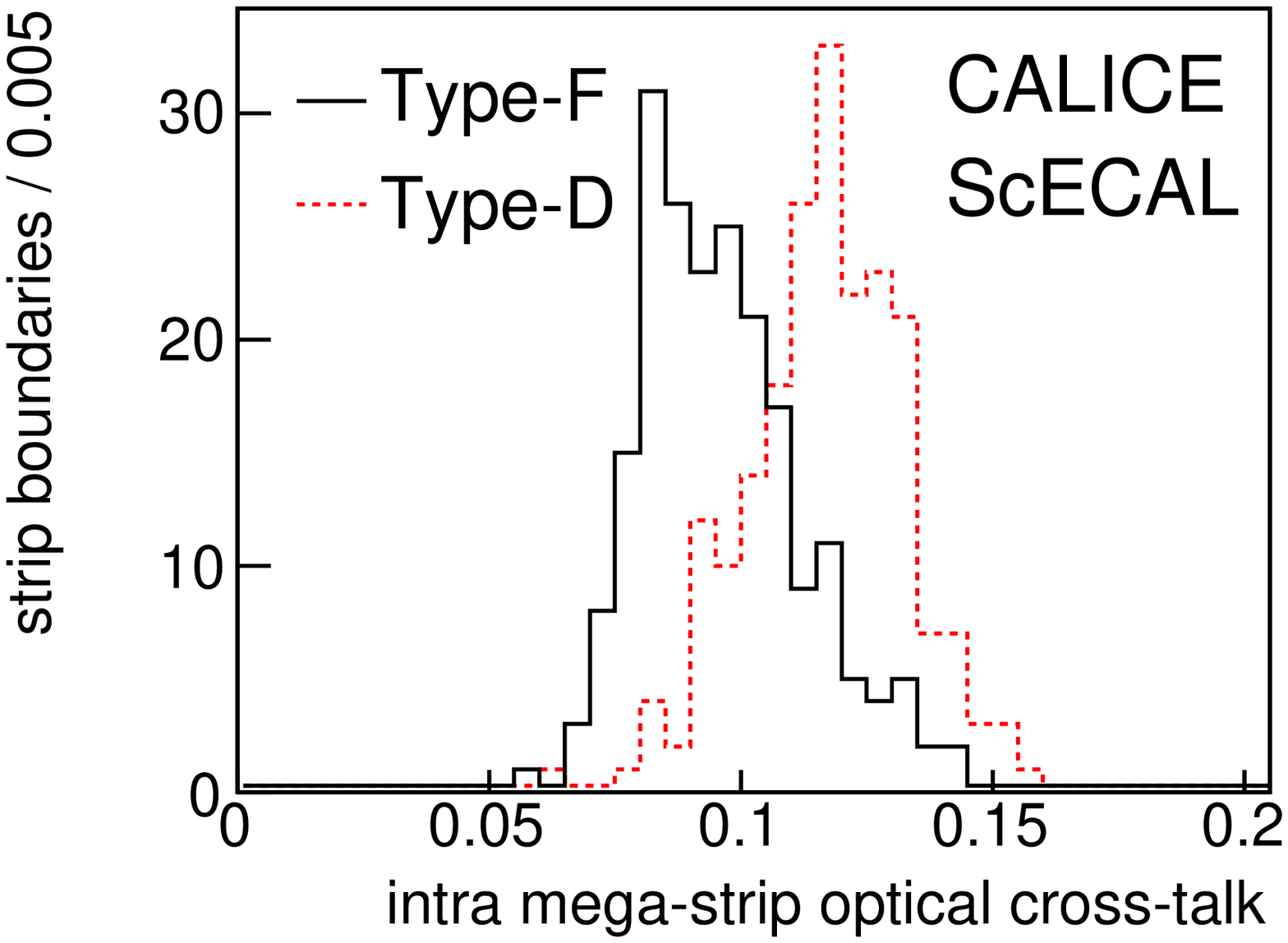}
}
\caption{
\textbf{Left:} Example of MPPC signal when a positron traversed a strip, a neighbouring strip within the same mega-strip or a non-neighbouring (other) strip. 
\textbf{Right:} Distribution of measured optical cross-talk between neighbouring strips in the same mega-strip.}
\label{fig:xtalk}
\end{figure}

Adjacent strips in the same mega-strip were not perfectly
optically isolated by the PET film inserted into the pairs of grooves (shown in Fig.~\ref{fig:mega-strip}).
The resulting optical cross-talk was measured using the calibration data by comparing the signals
in a given strip when the positron passed through the strip itself, an adjacent strip, 
or a more distant strip in the same mega-strip. It was measured separately across each strip boundary.
Figure ~\ref{fig:xtalk} shows an example of these signals in a particular strip, and
the distribution of the cross-talk measured between all pairs of strips within the same mega-strip.
The cross-talk between neighbouring strips is typically around 10\%, with a relative variation of around 15\% (RMS).
We observe that the cross-talk in type-D mega-strips is on average around \
21\% % jose claims 25%, let's check- difference in means is 21.45
larger than in type-F.
% , probably due to a higher collection efficiency for light leaking across strip boundaries.
The cross-talk between mega-strips in the same layer was also considered; it was measured to be smaller
than within the mega-strip, around 4\% on average. 

\begin{figure}[t]
\center{
\includegraphics[width=0.6\textwidth]{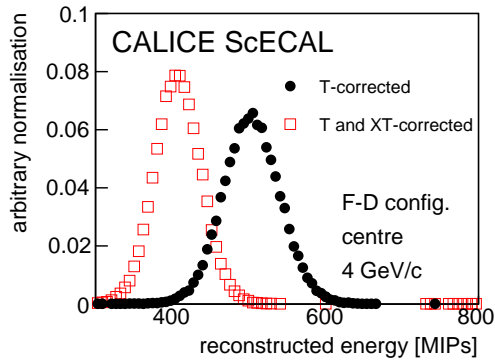}
}
\caption{
The reconstructed energy after the temperature correction (closed circles) and after an additional cross-talk (XT) correction (open squares), 
for the same sample of central events collected at 4~GeV/c by the \FD\ detector configuration.
}
\label{fig:xtalkcorrdemo}
\end{figure}

Since the MIP calibration was defined without accounting for cross-talk, 
a simple sum over measured strip energies would give an overestimate of the deposited energy
in terms of MIPs. If the cross-talk between strips is not uniform in the detector, energy deposited in
strips with larger cross-talk would get a larger weight in the sum. 
This is in effect a miscalibration, and would introduce an additional constant term into the energy resolution.
A cross-talk correction procedure was developed to subtract the estimated cross-talk contribution from the signal measured
in each strip.
The cross-talk between strips within a given layer can be expressed as a matrix ${\mathbf{M}_\mathrm{x-t}}$, which 
transforms the light produced in the strips $\mathbf{P}$ to the observed light distribution after cross-talk $\mathbf{O}$: 
$\mathbf{O} = \mathbf{M}_\mathrm{x-t} \cdot \mathbf{P}$. 
$\mathbf{M}_\mathrm{x-t}$ was defined to have entries of unity on the diagonal and the measured cross-talk between nearest
neighbours in the appropriate off-diagonal elements.
Since $\mathbf{M}_\mathrm{x-t}$ is almost diagonal, it is easily inverted and 
can then be used to subtract the effects of cross-talk: $\mathbf{P} = \mathbf{M}_\mathrm{x-t}^{-1} \mathbf{O}$. This
matrix was estimated separately for each detector layer using MIP calibration data, and applied to the EM shower data.
The effect of applying this cross-talk correction to EM shower data is shown in Fig.~\ref{fig:xtalkcorrdemo}.
The total energy is reduced by around 20\%, corresponding to the $\sim10\%$ leakage to the two 
neighbouring strips within the same mega-strip.

\subsubsection*{Estimation of systematic effects}
The difference in cross-talk between strips in the two mega-strip types is similar to the variation of cross-talk for strips of each type,
so a common systematic uncertainty was assigned. 
The total spread in measured cross-talk values around the mean is of order 50\%.
To estimate the systematic uncertainty due to this cross-talk correction procedure, 
results obtained with and without the application of the correction were compared.
Half of the difference between these two approaches 
(corresponding to the overall spread of cross-talk measurements) 
was assigned as a systematic uncertainty.

\subsection{Energy linearity and resolution}\label{sec:reslin}

EM shower data, taken with absorber plates installed between the scintillator layers, were collected using
beam momenta of 1 -- 6~GeV/c.
Events were pre-selected by requiring that the signals in the upstream trigger and veto
counters were consistent with the passage of a single positron: the signals in the two trigger counters 
were required to be consistent with a single MIP signal, 
while in the veto counter a pedestal-only signal was required.
The corrections described in previous sections were then applied to the data:
correction of the MPPC response non-linearity, conversion to MIP equivalent units
using the measured temperature-dependent calibration constants,
and the correction of optical cross-talk.
The measured event energy, in MIPs, was calculated as the sum of all strip energies.

The detector response was measured in the two configurations and in two detector regions. The ``central region'',
selected by requiring that the beam axis was centred within 1.5~mm of the detector centre $x, y=0$,
% when the shower centre-of-gravity in ($x, y$) was within XXcm \textbf{DANIEL: check!} of the detector centre ($x, y=0$), 
is the least affected by transverse shower leakage, 
% however most energy is deposited near the ends of the scintillator strips,
however most energy is deposited near the ends of the scintillator strips (near the boundary between megastrips),
making it the most susceptible to effects of any non-uniformity of the response along the strip length.
In the ``uniform region'', events in which the DCs reconstruct an impact on the front face of the ECAL within the 
four $10 \times 10~\mathrm{mm}^2$ areas centred at $x, y=\pm22.5~\mathrm{mm}$, 
particles pass near the strips' centres in both layer
orientations, minimising the effects of non-uniform strip response.

\begin{figure}[t]
\center{
\includegraphics[width=0.45\textwidth]{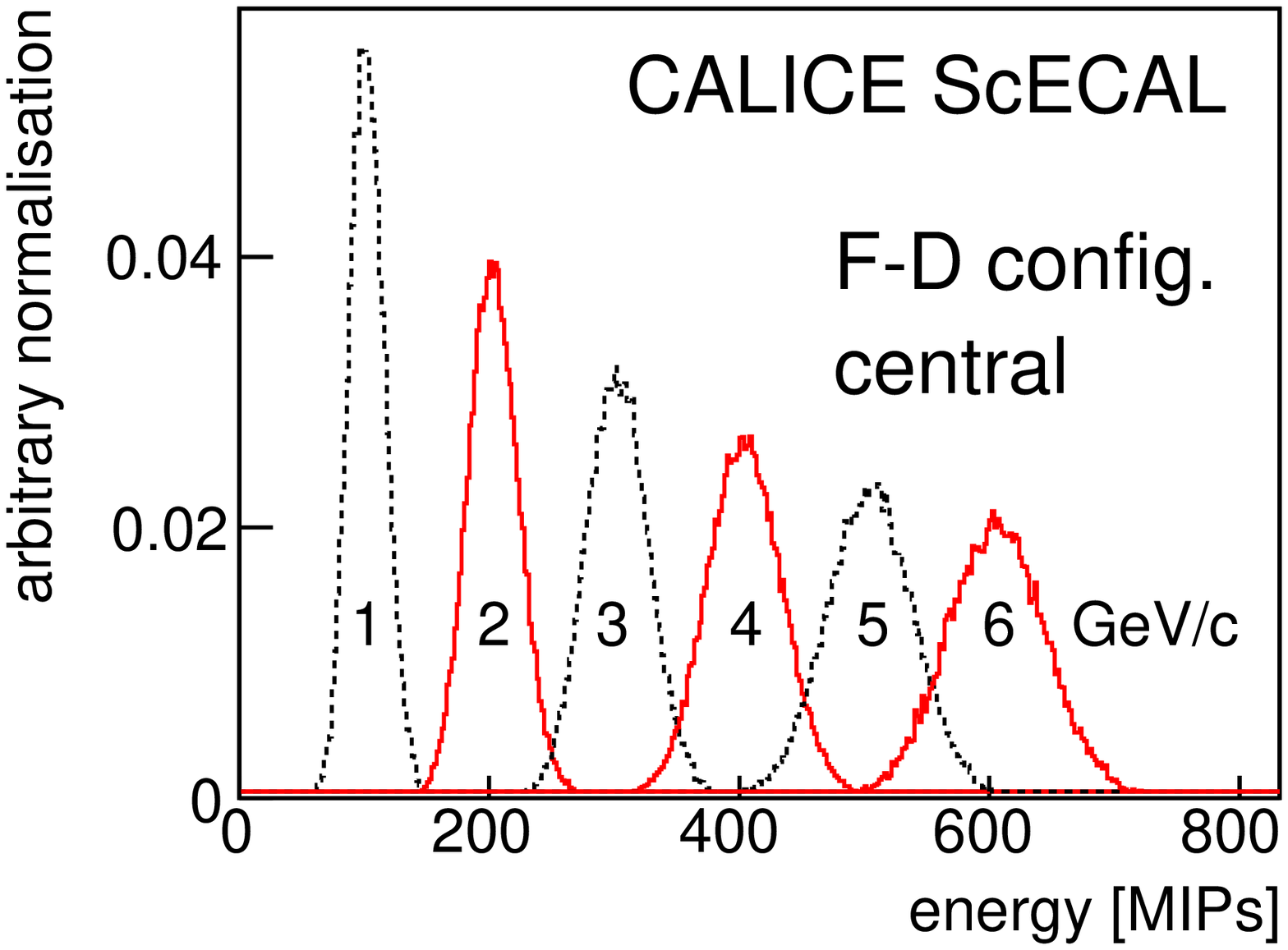}
\includegraphics[width=0.45\textwidth]{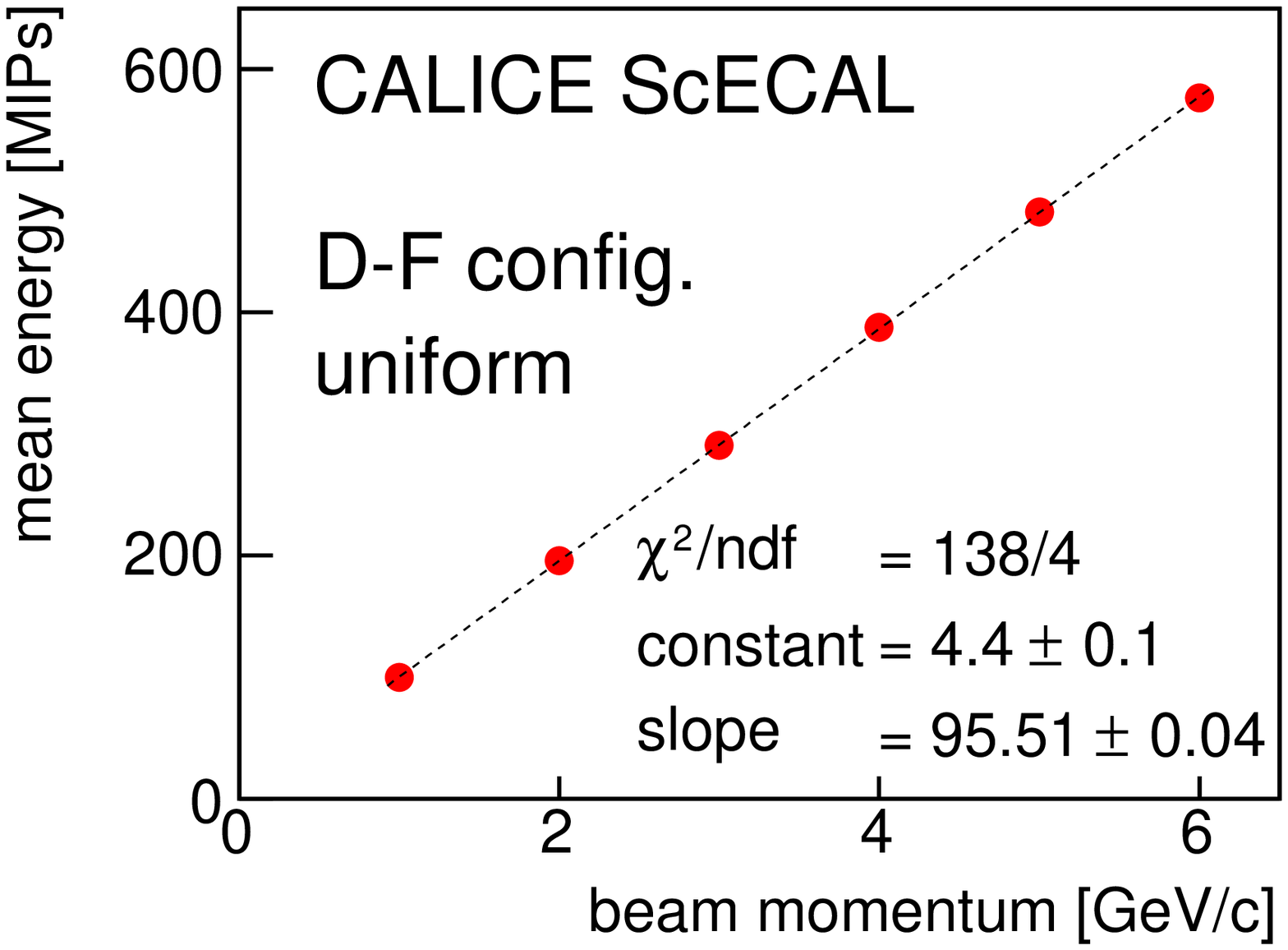}
}
\caption{
\textbf{Left:} The measured energy spectra of 1--6~GeV/c e$^+$ events collected in the central region, 
for the \FD\ detector configuration.
\textbf{Right:} The dependence of the measured mean energy response on the beam momentum in the uniform region, 
for the \DF\ detector configuration.
Only statistical uncertainties were used in the fit to the linear function (shown as a dotted line).
}
\label{fig:energyspectra}
\end{figure}

The distributions of the measured event energy 
for central events recorded by the \FD\ configuration
at each beam momentum are shown in
Fig.~\ref{fig:energyspectra}. 
Such distributions were made for both regions of both detector configurations, and
were fitted with Gaussian functions.
The dependence of the mean 
%$E$ % dj/jr 8/5/14
of the Gaussian on the beam momentum,
in the case of the uniform region of the \DF\ configuration,
is shown in the same figure.
%Fig.~\ref{fig:lin_final}.
% for the two 
%detector configurations, with the beam targeting the detector centre
%(``central region'') and the centre-strip region (``uniform region'').
These measurements were then fitted by a linear function.
The $\chi^2$ of the fit, in which systematic uncertainties were not considered, is rather high.
If, as an illustrative exercise, an uncertainty of 0.22\% is assigned to the average beam momentum, 
the $\chi^2$ per degree of freedom becomes exactly unity, 
and the constant and slope terms of the function change to 
$3.8\pm0.3$~MIP and $95.7\pm0.2$~MIP/(GeV/c), respectively.

\begin{figure}[t]
\center{
\includegraphics[width=0.45\textwidth]{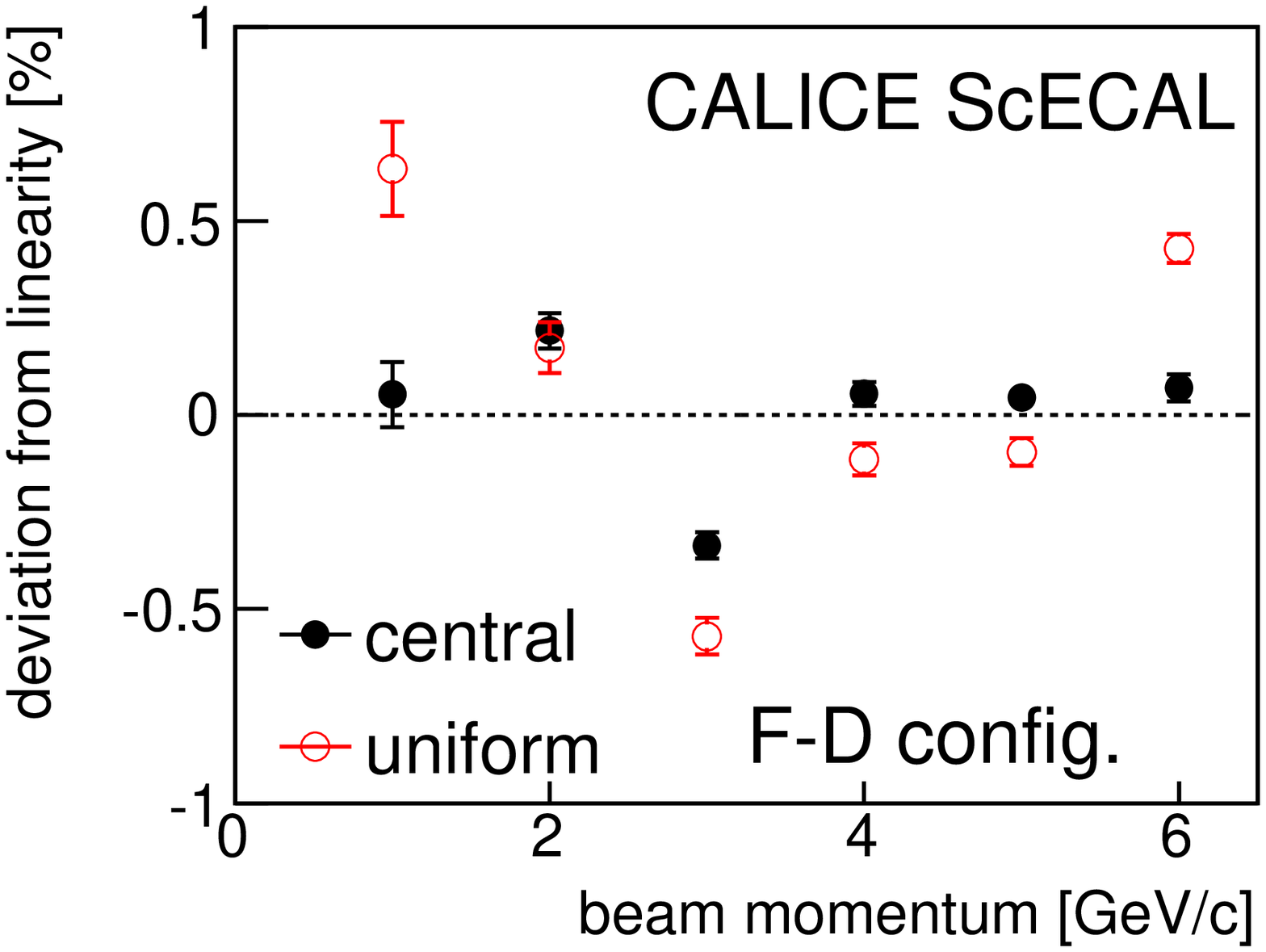}
\includegraphics[width=0.45\textwidth]{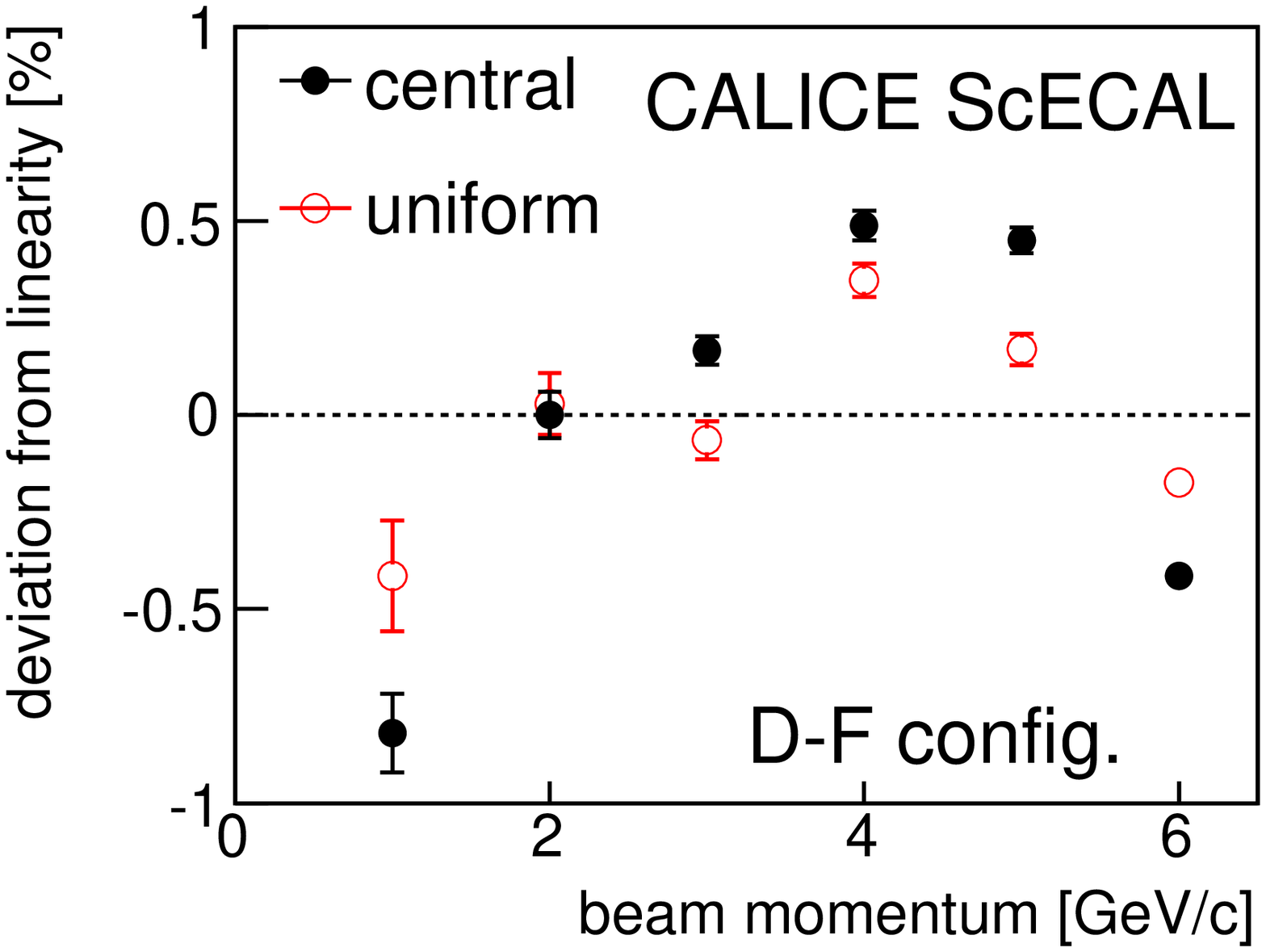}
}
\caption{Deviation from linear energy response measured in the central and uniform regions of both detector configurations.
Only statistical uncertainties are shown.}
\label{fig:energy_lin}
\end{figure}

The deviations from linear behaviour 
%, defined as
%\begin{equation}
%\mathrm{Deviation} = ( E_\mathrm{measured} - E_\mathrm{fit}) / E_\mathrm{beam},
%\end{equation}
$ ( E - E_\mathrm{fit}) / E_\mathrm{fit} $,
where $E_\mathrm{fit}$ is the prediction of the linearity fit,
%, and $E_\mathrm{beam}$ is the beam energy,
are shown in Fig.~\ref{fig:energy_lin}.
Deviations from linearity are within 1\% for all measured energies, configurations and detector regions.

\begin{figure}[t]
\center{
\includegraphics[width=0.45\textwidth]{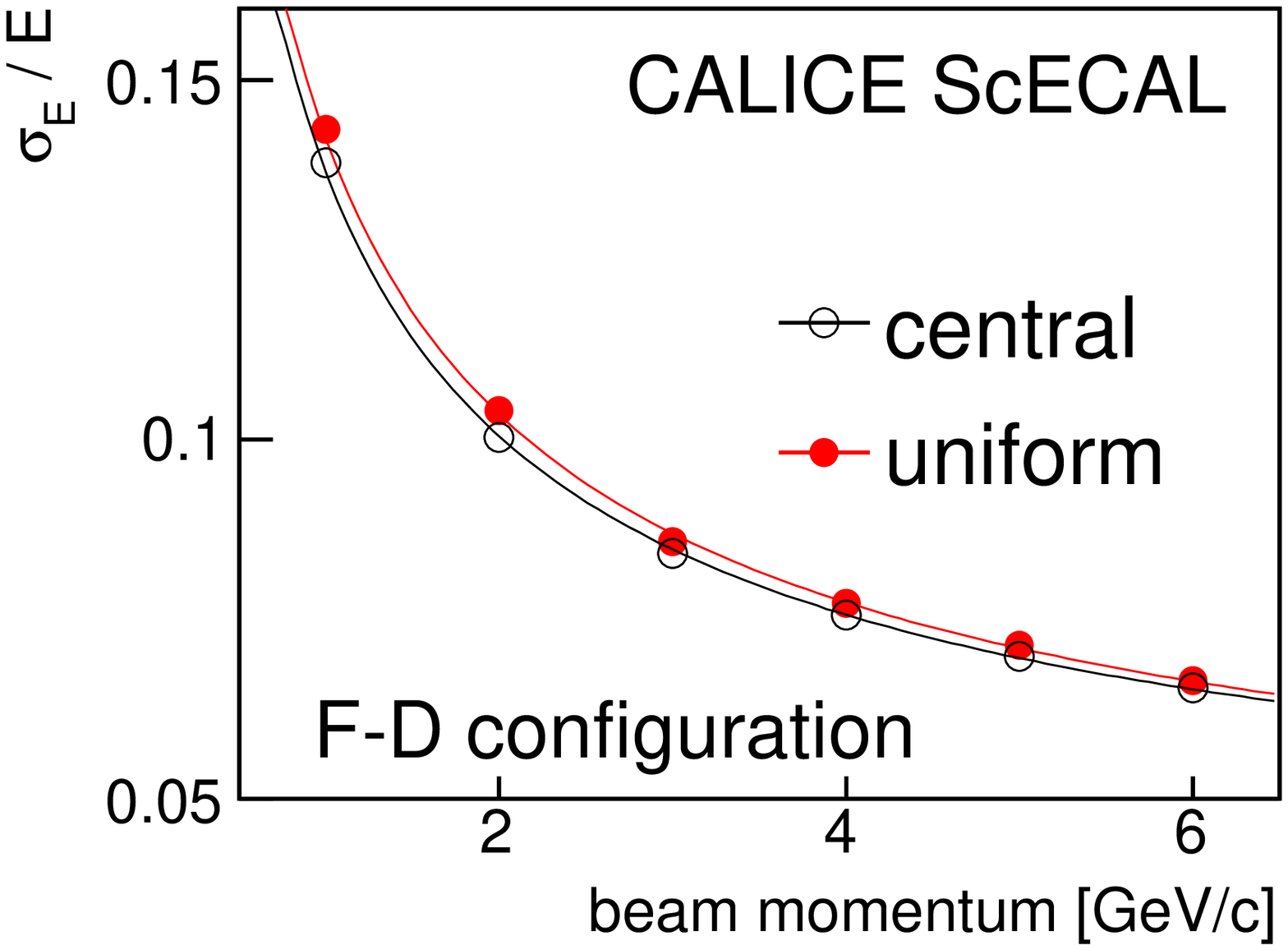}
\includegraphics[width=0.45\textwidth]{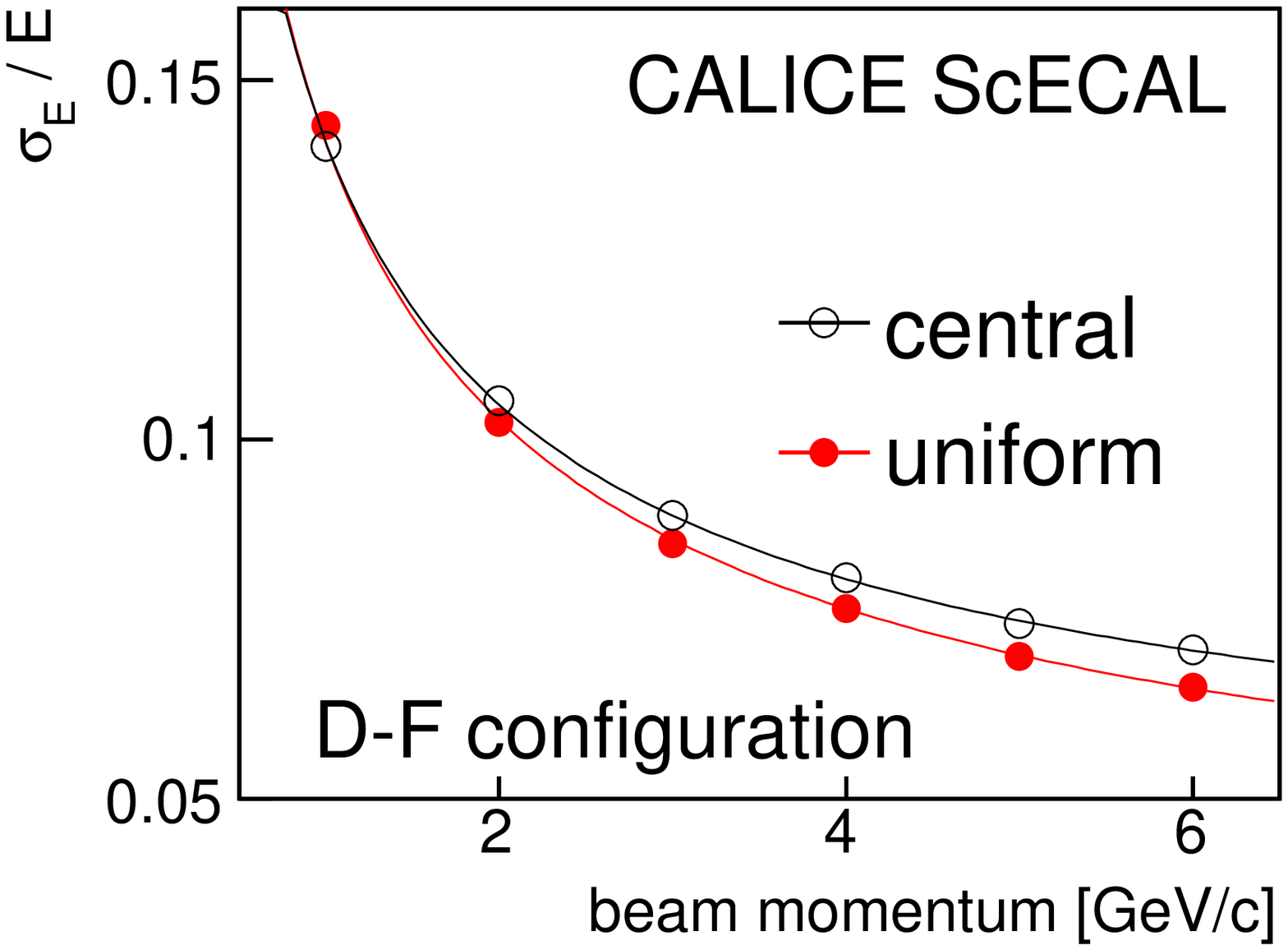}
}
\caption{The energy resolutions measured using data taken with 1--6~GeV/c e$^+$ beams in the central and uniform regions
of the two detector configurations. The results of the fits described in the text are shown, and
the fitted parameters reported in Table~\ref{tab:res_sys_grandsummary}.}
\label{fig:energy_resol}
\end{figure}

The relative width of the Gaussian (its width $\sigma_E$ divided by its mean $E$), was used to
estimate the energy resolution. Figure~\ref{fig:energy_resol} shows this relative width
as a function of the beam momentum in the two regions of the two detector configurations.
These points were fitted by a function of the form
\begin{equation}
 \frac{\sigma_E}{E} = \sqrt{ \frac{\sigma_{\mathrm{stochastic}}^{2}}{E_{\mathrm{beam}}\mathrm{[GeV]}} + \sigma_{\mathrm{constant}}^2 } \ ,\nonumber
\end{equation}
where $\sigma_{\mathrm{stochastic}}$ and $\sigma_{\mathrm{constant}}$ are
the stochastic and constant terms of the energy resolution, and $E_{\mathrm{beam}}$ is the beam energy in GeV.
The results of these fits are shown in the same figure, and are presented in Table~\ref{tab:res_sys_grandsummary}.
When a term describing detector noise was included in the above expression, its fitted value was always consistent with zero.

\begin{table}[t]
\begin{center} 
\begin{tabular}{c|l|l|ccc}\hline
configuration & region & & (\%) & statistical & systematic\\
\hline
\FD   & central & stochastic & 13.24 & $\pm 0.05$ & $\pm 0.20$ $^{+0}_{-1.66}$ \\
      &         & constant   &  3.65 & $\pm 0.05$ & $\pm 0.47$ $^{+0}_{-3.65}$\\
\cline{2-6}
      & uniform & stochastic & 13.76 & $\pm 0.07$ & $\pm 0.21$ $^{+0}_{-1.86}$ \\
      &         & constant   &  3.52 & $\pm 0.07$ & $\pm 0.47$ $^{+0}_{-3.52}$ \\
\hline
\DF   & central & stochastic & 13.43 & $\pm 0.06$ & $\pm 0.07$ $^{+0}_{-0.80}$ \\
      &         & constant   &  4.45 & $\pm 0.04$ & $\pm 0.22$ $^{+0}_{-4.45}$ \\
\cline{2-6}
      & uniform & stochastic & 13.73 & $\pm 0.08$ & $\pm 0.07$ $^{+0}_{-2.34}$ \\
      &         & constant   &  3.35 & $\pm 0.07$ & $\pm 0.22$ $^{+0}_{-3.35}$ \\
\hline
\end{tabular}
\end{center} 
\caption{Measured stochastic and constant terms of the \mbox{ScECAL} energy resolution in the 
central and uniform regions of the two detector configurations.
The second systematic uncertainty is the contribution due to the 
assumed 5\% beam energy spread.
}
\label{tab:res_sys_grandsummary}
\end{table}

\subsection{Summary of the systematic uncertainties}

The approach taken to the systematic uncertainties arising from the MIP calibration, saturation, 
temperature and cross-talk corrections are described in previous sections.
The observed shifts in the stochastic and constant terms of the energy resolution when
these procedures were applied were used as systematic uncertainties.
Where uniform systematic uncertainties were assumed (for example, in the cross-talk correction),
the average of the shifts in different configurations and regions was used.
The resulting uncertainties on the stochastic and constant terms of the energy resolution
are shown in Table~\ref{tab:res_sys_summary}. 

%table moved for ref#2
\begin{table}
\begin{center}
\begin{tabular}{l|c|l|c|c}\hline
 source & configuration & region & $\delta\sigma_\mathrm{stochastic}$ & $\delta\sigma_\mathrm{constant}$ \\
\hline
MIP calibration 
 & \FD  & central   & $\pm 0.01$ & $\pm 0.02$ \\
 &      & uniform   & $\pm 0.02$ & $\pm 0.02$ \\
\cline{2-5}
 & \DF  & central  & $\pm 0.01$ & $\pm 0.02$ \\
 &      & uniform  & $\pm 0.02$ & $\pm 0.02$ \\
\hline
Temperature correction
 & \FD  & central   & $\pm 0.02$ & $\pm 0.01$ \\
 &      & uniform   & $\pm 0.04$ & $\pm 0.02$ \\
\cline{2-5}
 & \DF  & central  & $\pm 0.01$ & $\pm 0.02$ \\
 &      & uniform  & $\pm 0.02$ & $\pm 0.02$ \\
\hline
Cross-talk correction
 & both & both & $\pm 0.03$ & $\pm 0.12$  \\
\hline
Single pixel signal %($d_\mathrm{low-gain}$)
 & both & both & $\pm 0.06$ &  $\pm 0.17$ \\
\hline
Effective pixel number 
 & \FD  & both & $\pm 0.19$ & $\pm 0.42$ \\ % this is case E
 & \DF  & both & $\pm 0.01$ & $\pm 0.07$ \\
\hline
\hline
Total
 & \FD                & central  & $\pm 0.20$ & $\pm 0.47$\\
(not including &      & uniform  & $\pm 0.21$ & $\pm 0.47$ \\
\cline{2-5}
beam energy spread) & \DF  & central  & $\pm 0.07$ & $\pm 0.22$\\
                    &      & uniform  & $\pm 0.07$ & $\pm 0.22$\\
\hline
\hline
Beam energy spread
 & \FD  & central  & $^{+0}_{-1.66}$ & $^{+0}_{-3.65}$ \\
(assumed to be 5\%)
 &      & uniform  & $^{+0}_{-1.86}$ & $^{+0}_{-3.52}$ \\
\cline{2-5}
 & \DF  & central  & $^{+0}_{-0.80}$ & $^{+0}_{-4.45}$ \\
 &      & uniform  & $^{+0}_{-2.34}$ & $^{+0}_{-3.35}$ \\
\hline
\end{tabular}
\end{center}
\caption{Summary of the systematic uncertainties on the
stochastic and constant terms of the energy resolution measurement.
The quoted figures are the absolute uncertainties on the stochastic and constant terms (in \%).
}
\label{tab:res_sys_summary}
\end{table}

The systematic uncertainties are dominated by uncertainties in the MPPC saturation correction,
both from the effective number of pixels which characterises the MPPC response enhancement and
the signal corresponding to a single fired pixel.
The uncertainty on the effective number of pixels stems from the fact that only a single
strip--MPPC unit was tested, necessitating a conservative estimate of the systematic uncertainty.
The uncertainty on the single pixel signal is dominated by the small fraction ($\sim 6\%$) of electronics channels
in which the ratio between the high and low gain modes was measured.

Any spread in the beam momentum will contribute to the width of the reconstructed energy distributions
shown in Fig.~\ref{fig:energyspectra}~(left). If this beam momentum spread is well understood,
it can be subtracted from the measured widths to give the intrinsic calorimeter
energy resolution.
However, the momentum spread of the test beam at DESY is not very well understood at present. 
An estimated upper limit on the momentum spread is given as $5\%$~\cite{DESY_testbeam}, but
the true spread is likely to be smaller, and may depend on energy\footnote{Norbert~Meyners (DESY), personal communication (2013).}.
Due to this uncertainty, we have chosen not to subtract 
a beam momentum spread for the nominal measurement, but assign a systematic uncertainty due 
to this effect. We estimate this uncertainty by subtracting, in quadrature, $5\%$ from the energy resolution
measured at each beam momentum. This has a large effect on the measured energy resolutions, 
reducing the fitted constant term to zero in each case, and also leading to significant reductions
of the stochastic terms. 
%This leads us to suspect that the true momentum spread is significantly smaller than 
%this 5\% upper limit.

% \clearpage

\section{Simulation}\label{sec:simulation}

A GEANT4 simulation of the prototype was developed using the Mokka package~\cite{mokka}.
Active scintillator layers were segmented into strips, 
with an insensitive region corresponding to the MPPC package at one end of each strip.
The simulated module was significantly larger than the prototype in both the transverse and longitudinal dimensions,
eliminating the effects of energy leakage;
to estimate the effect of leakage, only energy deposited within a volume corresponding
to the prototype was considered.
Type-D (-F) strips were modeled as having a non-uniform response along their length, 
characterised by an exponential function with an attenuation length of 
% 150~mm (300~mm), consistent with the measured non-uniformity.
115~mm (280~mm), the average of several measurements of strip uniformity made
using data collected in detailed position scans performed during the beam test.
The root mean square variation of the measured attenuation lengths was used to set a 
systematic uncertainty on the simulated results.

The energy response of the model to single positrons in the range 1 to 6~GeV was estimated for a number of different scenarios,
to identify impacts of the detector characteristics on the response: \\
%\hspace*{0.3cm} 
a) large detector size (no energy leakage), no insensitive MPPC volume, and uniform strip response;\\
%\hspace*{0.3cm} 
b) same as (a), but with insensitive MPPC volume;\\
%\hspace*{0.3cm} 
c) same as (b), but with same size as prototype;\\
%\hspace*{0.3cm} 
d) same as (c), but with non-uniform strip response (\FD\ configuration);\\
%\hspace*{0.3cm} 
e) same as (c), but with non-uniform strip response (\DF\ configuration).

The energy resolution was simulated in the ``central'' and ``uniform'' regions of each of these scenarios.
% are summarised in 
%Table~\ref{tab:simulation_results}.
These simulations were used to estimate the contributions to the constant term from the
MPPC volume, limited prototype size, and strip non-uniformity, as shown in Table~\ref{tab:const_contribs_digest}. 
The contribution due to the non-uniformity is the largest in the ``central'' region, 
while the energy leakage is the most significant contribution in the ``uniform'' region. 

% \begin{table}[t]
% \begin{center}
% \begin{tabular}{l|c|c|c|l|c|c}\hline
%         & MPPC       & energy      & strip      &         & stochastic & constant \\
%         & vol        & leakage     & non-uni.   &         &  (\%)      & (\%) \\
% \hline
% \hline
% a &            &             &            & central & $13.19 \pm 0.09 $ & $ 1.56 \pm 0.19 $ \\
%   &            &             &            & uniform & $13.37 \pm 0.08 $ & $ 1.40 \pm 0.19 $ \\
% \hline
% b & \checkmark &             &            & central & $13.27 \pm 0.09 $ & $ 2.52 \pm 0.13 $ \\
%   &            &             &            & uniform & $13.18 \pm 0.08 $ & $ 1.68 \pm 0.16 $ \\
% \hline
% c & \checkmark &  \checkmark &            & central & $13.39 \pm 0.10 $ & $ 3.28 \pm 0.11 $ \\
%   &            &             &            & uniform & $13.54 \pm 0.08 $ & $ 2.33 \pm 0.13 $ \\
% \hline
% d & \checkmark & \checkmark  & \FD        & central & $13.24 \pm 0.10 $ & $ 4.26 \pm 0.09 $ \\
%   &            &             &            & uniform & $13.45 \pm 0.08 $ & $ 2.61 \pm 0.12 $ \\
% \hline
% e & \checkmark &  \checkmark & \DF        & central & $13.22 \pm 0.11 $ & $ 5.12 \pm 0.08 $ \\
%   &            &             &            & uniform & $13.52 \pm 0.09 $ & $ 2.57 \pm 0.12 $ \\
% \hline
% 
% \hline
% \end{tabular}
% \end{center}
% \caption{{\bf TEMPORARY, for information only. I don't propose to give whole table in the paper.}
% Energy resolutions predicted by Monte Carlo simulation in the various scenarios described in the text.
% }
% \label{tab:simulation_results}
% \end{table}

\begin{table}[t]
\begin{center}
\begin{tabular}{l|c|c}
\hline
              & central       & uniform \\
\hline
MPPC volume   & $2.0 \pm 0.2 \%$ & $ 0.9 \pm 0.4 \%$ \\
leakage       & $2.1 \pm 0.2 \%$ & $ 1.6 \pm 0.2 \%$ \\
%non-uni (\FD) & $2.7 \pm 0.2 \%$ & $ 1.2 \pm 0.4 \%$ \\
%non-uni (\DF) & $3.9 \pm 0.1 \%$ & $ 1.1 \pm 0.4 \%$ \\
non-uni (\FD) & $3.3 \pm 0.2 \pm 0.9\%$ & $ 1.4 \pm 0.3 \pm 0.1\%$ \\
non-uni (\DF) & $4.9 \pm 0.1 \pm 1.1\%$ & $ 1.4 \pm 0.3 \pm 0.2\%$ \\
\hline
\end{tabular}
\end{center}
\caption{
Contributions to the constant term of the energy resolution in the two detector regions,
estimated by Monte Carlo simulations. In the case of the non-uniformity (``non-uni''), the first uncertainty
is statistical, the second due to variations of the strip attenuation length.
}
\label{tab:const_contribs_digest}
\end{table}

\begin{figure}[t]
\center{
\includegraphics[width=0.45\textwidth]{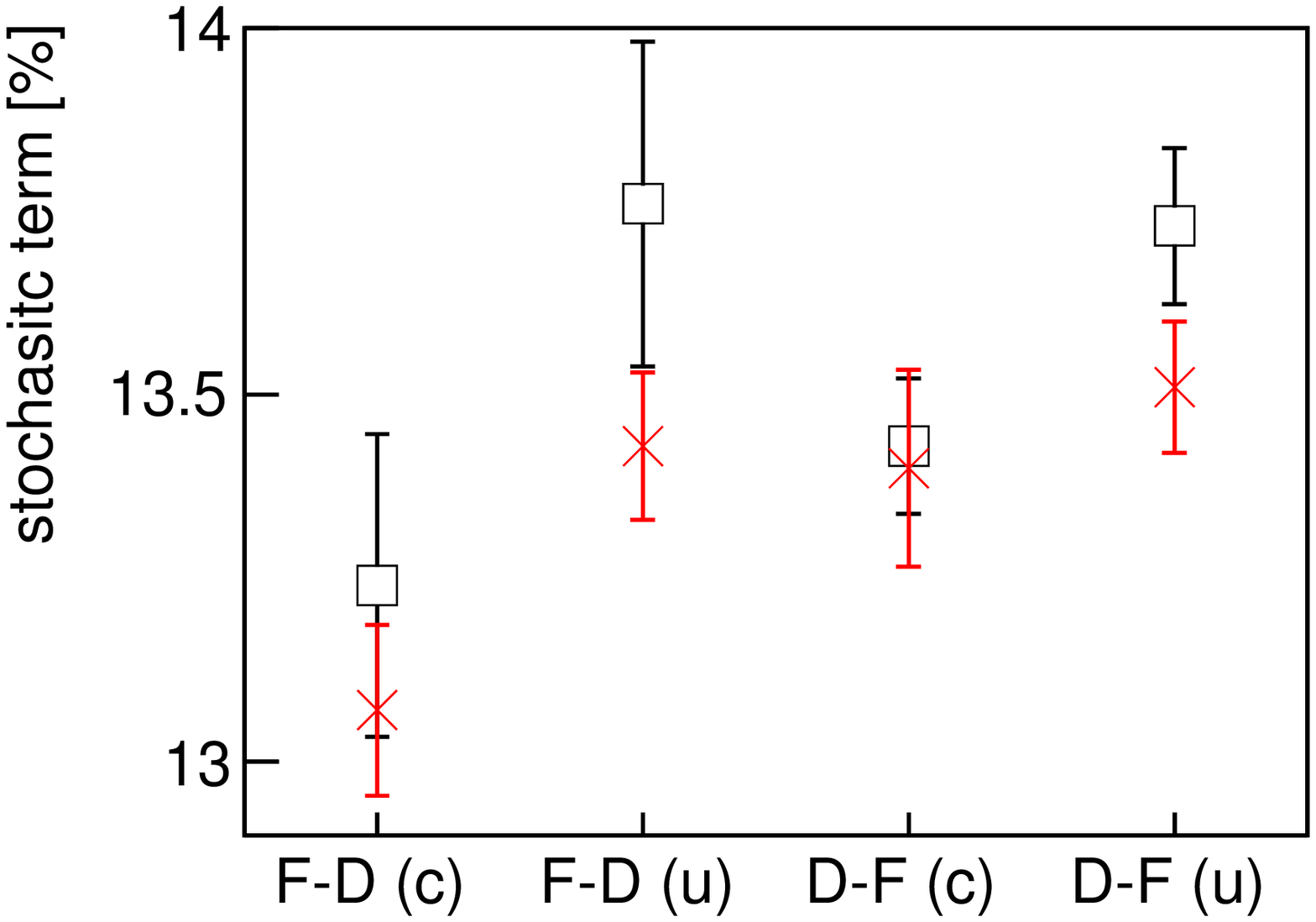}
\includegraphics[width=0.45\textwidth]{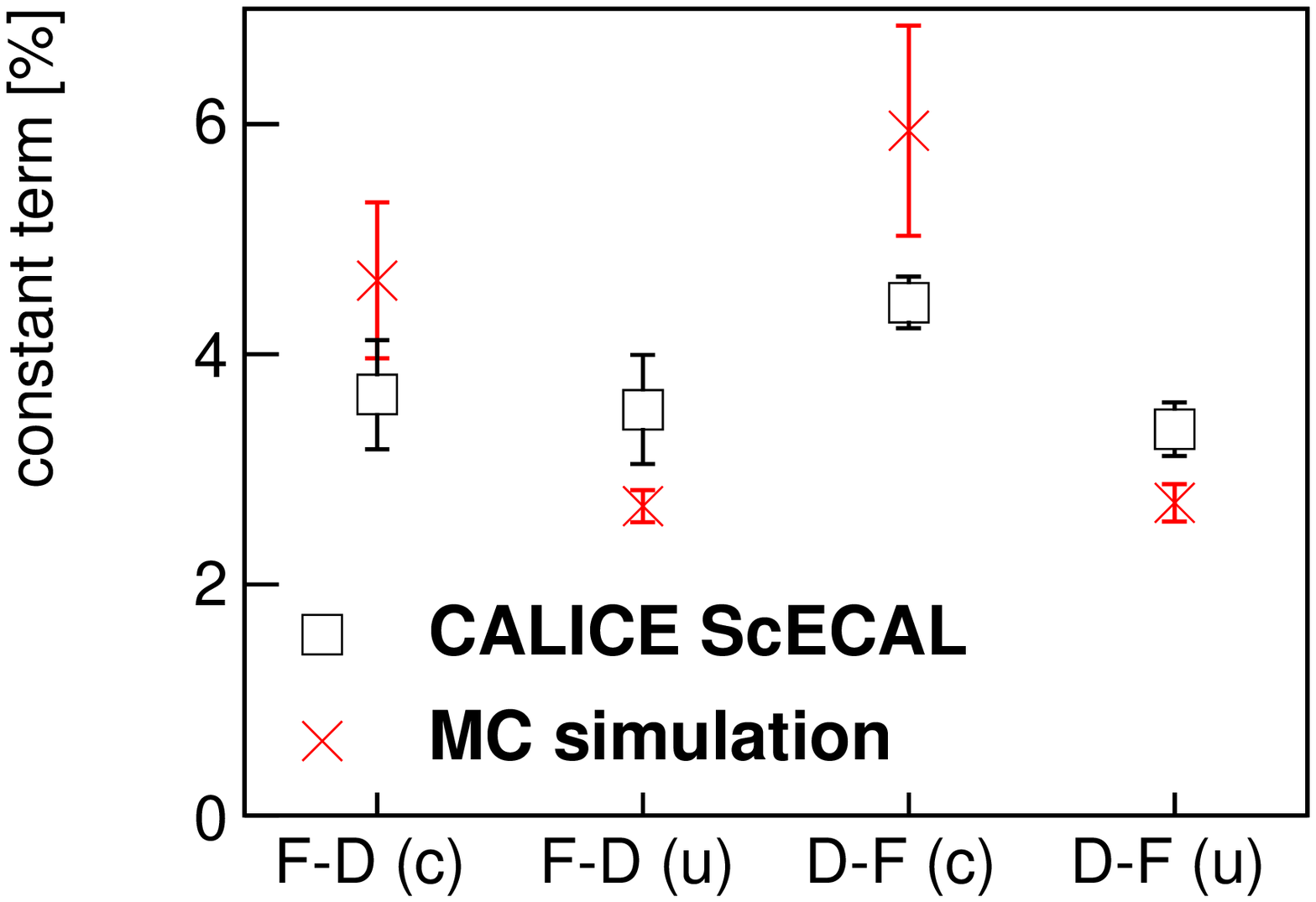}
}
\caption{
Comparison of the stochastic (left) and constant (right) terms of the energy resolution measured in data and simulations
in the central (``c'') and uniform (``u'') regions of both detector configurations.
The beam energy spread has neither been subtracted from the data, nor included in the simulations.
The error bars on the data correspond to the quadratic sum of the statistical uncertainty and all systematic uncertainties 
except that due to the beam energy spread. The error bars on the simulation points include the statistical uncertainty
and systematic effects related to the strip attenuation length.
}
\label{fig:sim_data_res}
\end{figure}

The predicted and measured energy resolutions in the different regions and configurations are compared in 
Fig.~\ref{fig:sim_data_res}. 
The stochastic term of the energy resolution measured in data is consistent with the Monte Carlo prediction in all 
detector configurations and regions.
% The agreement between the measured and predicted resolutions is reasonable.
The constant term measured in the uniform region is around 1\% larger than the predicted one in both detector configurations, possibly
due to the beam momentum spread, which is not accounted for in the simulation. The size of the discrepancy suggests that the true beam 
momentum spread is significantly smaller than the conservative bound of 5~\% quoted by the test beam operators.
In the central region, the measured constant term is consistently smaller than the simulated one, however this difference 
is similar in size to the systematic uncertainty due to the scintillator strip attenuation length. 
% This suggests that the non-uniformity of the scintillator response, modelled using a 
% simple exponential dependence along the strip length, may be weaker in the data than in the simulation.

% \clearpage

\section{Summary}\label{sec:summary}

During this test beam campaign, several hundred MPPCs were successfully operated in a prototype scintillator strip-based ECAL.
This demonstrates that such a technique is feasible and represents an important milestone in the development of a 
\mbox{ScECAL} for a future high energy lepton collider. The applied temperature-based corrections to the MPPC response successfully
stabilised the prototype's response.

The energy measurement of this calorimeter prototype was measured to be linear to within 1\% in the energy 
range between 1 and 6 GeV. 
% (in the absence of any strongly energy-dependent systematic uncertainties). <-- daniel no longer remembers what he means by this...
The residual non-linearities do not depend strongly on the detector configuration
or the region of the detector (central or uniform).

The stochastic terms in the various configurations and regions are measured to be between 13 and 14\%, 
sufficient for the use of Particle Flow Algorithms at future lepton collider detectors.
The constant term was measured to be between 3 and 4.5\%. 
% Depending on the true beam energy spread, the intrinsic calorimetric performance may be better than this.
% somewhat better than these numbers.
The energy resolution measured in the \FD\ configuration (whose performance is dominated by the type-F(ibre) module), 
is very similar in the two detector regions.
%is better in the central region than the uniform regions, probably due to the better lateral 
%energy containment in the centre. 
In the \DF\ configuration, the constant term is significantly larger in the 
central region. This difference is attributed to the effect of non-uniform response along the strip length, 
and the effect of the dead volume due to the MPPC package.
% in particular that due to the increased light collection efficiency close to the MPPC.
A simulation study supports these conclusions, and has shown that major contributions to the 
constant term of the energy resolution are 
the non-uniformity of the strip response, 
the limited size of the prototype,
and the insensitive volume due to the MPPC package.

% In general, the size of the measured constant term is rather large. We expect that with a larger detector 
% (to reduce shower leakage both laterally and longitudinally), this will be significantly reduced.
% A better knowledge and treatment of the beam energy spread will
% allow the intrinsic \mbox{ScECAL} resolution to be measured without this contribution.

% daniel changes 12/2/2014
The present study has highlighted various aspects of the \mbox{ScECAL} design
which must be addressed before building a full detector:
improved uniformity of the scintillator strip response, and
reduced dead volume due to the MPPC package (to reduce the constant term of the energy resolution);
development of an improved LED-based MPPC monitoring system able to monitor the gain of all channels; 
tests of a prototype with larger volume, to measure the constant term without leakage effects;
tests of this larger detector using electron and other particle beams with a larger energy range; and
the use of embedded front-end electronics, to realise the high channel density and small number of external cables
required for integration into a future collider experiment.
The results of tests of a larger ScECAL prototype, which address a number of these issues, will be reported in a later paper.

\section*{Acknowledgements}

We gratefully acknowledge the DESY management for its support and
hospitality, and the DESY accelerator staff for the reliable and efficient
beam operation. 
We would like to thank the HEP group of the University of Tsukuba for the loan of drift chambers.
This work was supported by the Bundesministerium f\"{u}r Bildung und Forschung, Germany; 
by the  the DFG cluster of excellence `Origin and Structure of the Universe' of Germany;  
by the Helmholtz-Nachwuchsgruppen grant VH-NG-206; 
by the BMBF, grant no. 05HS6VH1; 
by JSPS KAKENHI Grant-in-Aid for Scientific Research numbers 17340071 and 18GS0202;
by the Russian Ministry of Education and Science contracts 8174, 8411,
1366.2012.2, and 14.A12.31.0006;
by MICINN and CPAN, Spain;
by CRI(MST) of MOST/KOSEF in Korea; 
by the US Department of Energy and the US National Science Foundation; 
by the Ministry of Education, Youth and Sports of the Czech Republic
under the projects AV0 Z3407391, AV0 Z10100502, LC527  and LA09042 and by the
Grant Agency of the Czech Republic under the project 202/05/0653;  
by the National Sciences and Engineering Research Council of Canada; 
and by the Science and Technology Facilities Council, UK.

\bibliographystyle{elsarticle-num} % NIM

\bibliography{DESYpaper}

\end{document}